\documentclass[twocolumn, fleqn]{article}

\usepackage[svgnames, x11names]{xcolor}
\usepackage[english]{babel}
\usepackage[utf8]{inputenc}

\usepackage{blindtext}
\usepackage{lipsum}
\usepackage{url}
\usepackage{listings}
\usepackage{graphicx}

\usepackage{tikz}
\usetikzlibrary{spy, shapes, backgrounds, fit, angles, quotes, arrows, math, decorations.markings, arrows.meta, decorations.pathmorphing, patterns, fadings, hobby}

\usepackage{pgfplots}
\pgfdeclarelayer{foreground}
\pgfsetlayers{background,main,foreground}
\DeclareUnicodeCharacter{2212}{−}
\usepgfplotslibrary{groupplots, dateplot, fillbetween}

\usepackage{tcolorbox}
\usepackage{soul}

\usepackage{float}
\usepackage{xcolor}
\usepackage{caption}
\usepackage{subcaption}
\usepackage{xpatch}
\usepackage{appendix}
\usepackage{amsmath,amsfonts,amssymb}
\usepackage{bm}
\usepackage{gensymb}
\usepackage{overpic}

\usepackage{enumitem}

\usepackage{pgfplots}
\pgfplotsset{width=10cm,compat=1.9}
\usepgfplotslibrary{external}

\usepackage{fancyvrb} 

\usepackage{scalerel}
\usepackage{siunitx}
\newcommand{\unit}[1]{\si[sticky-per]{#1}}

\makeatletter
\let\c@author\relax
\makeatother

\usepackage[backend=biber, style=numeric-comp, sorting=none, maxbibnames=10]{biblatex}

\usepackage{csquotes}
\usepackage{pifont}

\usepackage[%
	paperwidth=210mm,
	paperheight=280mm,
	vmargin={19.5mm,18.2mm},
	hmargin={15mm,15mm},
	headsep=12pt,
	footskip=12pt,
	columnsep=18pt
]{geometry}

\usepackage{multirow}

\usepackage{hyperref}

\definecolor{backcolor}{RGB}{240, 240, 240}

\makeatletter
\newlength{\fboxrsep}
\newlength{\fboxlsep}
\newlength{\fboxtsep}
\renewcommand\arraystretch{1.5}

\newcommand{\density}{\unit{\kilo\gram\per\cubic\metre}}

\newcommand{\K}{\unit{\kelvin}}
\newcommand{\mK}{\unit{\milli\kelvin}}
\newcommand{\uA}{\unit{\micro\ampere}}
\newcommand{\Amp}{\unit{\ampere}}
\newcommand{\uF}{\unit{\micro\farad}}
\newcommand{\mJ}{\unit{\milli\joule\per\cubic\centi\metre}}
\newcommand{\kW}{\unit{\kilo\watt\per\square\metre}}

\newcommand{\kOhm}{\unit{\kilo\ohm}}

\newcommand{\partialt}[1]{\ensuremath{\frac{\partial #1}{\partial t}}}
\newcommand{\partialx}[1]{\ensuremath{\frac{\partial #1}{\partial x}}}

\DeclareMathAlphabet\mathbfcal{OMS}{cmsy}{b}{n}

\newcommand{\ecco}{Eccobond\textsuperscript{\textregistered}}
\newcommand{\cx}{Cernox\textsuperscript{\textregistered}}

\newcommand{\Ts}{\ensuremath{T_\text{s}}}
\newcommand{\Tb}{\ensuremath{T_\text{b}}}

\newcommand{\Tl}{\ensuremath{T_\lambda}}

\newcommand{\QK}{\ensuremath{Q_\text{K}}}
\newcommand{\aK}{\ensuremath{a_\text{K}}}
\newcommand{\nK}{\ensuremath{n_\text{K}}}

\newcommand{\Qapp}{\ensuremath{Q_\text{app}}}

\newcommand{\aKUnit}{\unit{\watt\per\square\metre\kelvin\tothe{\nK}}}

\newcommand{\etal}{\emph{et al.}}

\newcommand{\us}{\unit{\micro\second}}
\newcommand{\ms}{\unit{\milli\second}}

\newcommand{\um}{\unit{\micro\metre}}

\newcommand{\df}[2]{\ensuremath{\frac{\text{d} #1}{\text{d} #2}}}

\newcommand{\sensor}[1]{%
	\mbox{%
		\vphantom{#1}\smash{%
			\tcbox[
				on line,
				boxsep=0pt,
				left=2.5pt, right=2.5pt,
				top=2pt, bottom=2pt,
				colback=LightGray, colframe=White,
				boxrule=0pt,
				arc=1mm, auto outer arc,
			]{\texttt{\textbf{#1}}}
		}%
	}%
	\hspace{-7pt}
}

\makeatletter
\patchcmd{\SOUL@ulunderline}{\dimen@}{\SOUL@dimen}{}{}
\patchcmd{\SOUL@ulunderline}{\dimen@}{\SOUL@dimen}{}{}
\patchcmd{\SOUL@ulunderline}{\dimen@}{\SOUL@dimen}{}{}
\newdimen\SOUL@dimen
\makeatother

\DeclareRobustCommand{\legendEntry}[1]{%
	{\sethlcolor{LightSteelBlue!35!}%
		\hl{\,#1\,}%
	}%
}

\tikzset{
	pointer/.style={
		thick,
		shorten >= 4pt,
		shorten <= 0pt,
		decoration={markings, mark={at position 1 with {\arrow{latex[line width=0.4pt, length=2.5pt,width=4pt]}}}},
		postaction=decorate
	},
	arrow/.style={
		thick,
		shorten >= #1,
		shorten <= 2pt,
		decoration={markings, mark={at position 1 with {\arrow{latex[line width=0.4pt, length=2.5pt,width=4pt]}}}},
		postaction=decorate
	},
	arrow/.default=2pt,
	arrowreversed/.style={
		thick,
		shorten >= #1,
		shorten <= 2pt,
		decoration={markings, mark={at position 0 with {\arrow{latex[line width=0.4pt, length=2.5pt,width=4pt]}}}},
		postaction=decorate
		 },
	arrowreversed/.default=2pt,
	dot/.style={
		thick,
		shorten >= 2pt,
		shorten <= 2pt,
		decoration={
			markings, mark={at position 1 with {\draw circle [radius=2pt];}}
		},
		postaction=decorate
	},
	dimen/.style={
		thick,
		shorten >= 4pt,
		shorten <= 4pt,
		decoration={
			markings, 
			mark = {
				at position 0 with {
					\arrowreversed{latex[line width=0.4pt, length=2.5pt,width=4pt]|[width=5mm]}
				}
			},
			mark = {
				at position 1 with {
					\arrow{latex[line width=0.4pt, length=2.5pt,width=4pt]|[width=5mm]}
				}
			}
		},
		postaction=decorate
	},
	dimenShort/.style={
		thick,
		shorten >= 4pt,
		shorten <= 4pt,
		decoration={
			markings, 
			mark = {
				at position 0 with {
					\arrowreversed{latex[line width=0.4pt, length=2.5pt,width=4pt]|[width=5mm]}
				}
			},
		},
		postaction=decorate
	},
	declare function={
		atan3(\a,\b)=ifthenelse(atan2(0,1)==90, atan2(\a,\b), atan2(\b,\a));
	},
	kinky cross radius/.initial=+.125cm,
	@kinky cross/.initial=+,
	kinky crosses/.is choice,
	kinky crosses/left/.style={@kinky cross=-},
	kinky crosses/right/.style={@kinky cross=+},
	kinky cross/.style args={(#1)--(#2)}{
		to path={
			let \p{@kc@}=($(\tikztotarget)-(\tikztostart)$),
			\n{@kc@}={atan3(\p{@kc@})+180} in
			-- ($(intersection of \tikztostart--{\tikztotarget} and #1--#2)!%
			\pgfkeysvalueof{/tikz/kinky cross radius}!(\tikztostart)$)
			arc [ radius     =\pgfkeysvalueof{/tikz/kinky cross radius},
			start angle=\n{@kc@},
			delta angle=\pgfkeysvalueof{/tikz/@kinky cross}180 ]
			-- (\tikztotarget)
		}
	}
}

\pgfplotsset{%
	grid style={
		line width=0.1pt, gray!50!
	},
	every axis/.append style={
		label style={font=\small},
		tick label style={font=\small}  
	}
}

\makeatletter
\renewcommand{\paragraph}{%
	\@startsection{paragraph}{4}%
	{\z@}{1mm \@plus 1ex \@minus .2ex}{-1em}%
	{\normalfont\normalsize\bfseries}%
}
\makeatother

\begin{document}	
	\title{Millisecond Time--Scale Measurements of\\Heat Transfer to an Open Bath of He II}
	
	\author{
		Jonas Blomberg Ghini$^{1,2,\ast}$ \and Bernhard Auchmann$^{2}$ \and Bertrand Baudouy$^{3}$
	}
	\date{%
		$^{1}$Department of Physics, Norwegian University of Science and Technology, NTNU, Norway\\%
		$^{2}$European Organization for Nuclear Research, CERN, Switzerland\\%
		$^{3}$Irfu, CEA, Université Paris--Saclay, F-91191 Gif--sur--Yvette, France\\%
		$^{\ast}$Corresponding author: \emph{jonas.blomberg.ghini@ntnu.no}\\%
	}
	\maketitle
	
	\begin{abstract}
		We explore steady state and transient heat transfer from a narrow, rectangular stainless steel heater strip cooled from one side by an open bath of He~II. Setup validation is done by fitting the Kapitza heat transfer expression $Q = \aK\,\left(\Ts^{\nK} - \Tb^{\nK}\right)$ to steady state measurements, finding fit parameters within the expected range; \aK\ = 1316.8$\pm10\%$~\aKUnit, \nK\ = 2.528$\pm10\%$. 

We find critical heat flux in line with estimates from literature, and the time between a step in heating and the onset of film boiling follows the expected $\propto Q^{-4}$ dependence. 

During the first millisecond after a step in applied heating power density our measurements show a slower thermal rise time than that found by a time--dependent one--dimensional model of our setup using the steady state Kapitza heat transfer expression as the cooling boundary condition. However, the results compare favourably with transient measurements in literature. After the first millisecond, agreement between measurement and model is excellent.

We do not find conclusive evidence of an orientation dependence of the Kapitza heat transfer mechanism, nor heat transfer differences that can be attributed to local surface variations along the same heater.

	\end{abstract}
	
	\section{Introduction}
		During operation of a particle accelerator, such as the Large Hadron Collider (LHC), it is inevitable that some particles from the circulating beam are lost, depositing their kinetic energy in equipment surrounding the beam pipe\cite{lossesAreInevitable}. There are three main sources of beam loss\cite[p. 370]{severalKindsOfBeamLosses}; 1) malfunctioning equipment, such as a magnet losing power, thus no longer bending the beam; 2) beam instabilities that over time cause parts of, or the whole, beam to veer off course; 3) scattering events in which some particles from the beam collide with stray matter in the beam pipe outside the dedicated interaction regions of the machine. Losses of the first and second kind usually arise at the collimators which intercept the stray beam\cite{collimators}, leading to particle showers absorbed in the machine components downstream\cite{lossesDownstreamOfCollimators}.

In the work presented here, the loss event of most relevance is characterised by the beam interacting with a dust particle in the beam pipe, which gives rise to a particle shower that deposits energy, over the course of about 1~ms, into the bulk of the superconducting magnets that surround the beam pipe along the length of the bending sections of the LHC\cite{ufoIS_and_hoursAfterQuench}. This kind of loss event is called a UFO event, as the dust particle is an Unidentified Falling Object. UFOs occur about 10 to 30 times per hour of operation of the LHC\cite[Fig. 2]{ufoIS_and_hoursAfterQuench}. If the energy deposition is sufficiently large, a UFO can quench a magnet, meaning the superconducting magnet, locally, becomes normal conducting\cite[p. 656]{iwasa_quenchIs}. In large accelerator magnets this transition is usually irreversible, and to protect the magnet from damage the beam is dumped, and the magnet quench protection system kicks in to dissipate the energy stored in the magnetic field. This aborts the operation of the LHC, and it then takes several hours to return the machine to normal operating conditions\cite{ufoIS_and_hoursAfterQuench}.

Analysing this kind of transient beam loss event, with the aim of determining whether or not the magnet would quench, requires modelling the physical behaviour of the system. This includes the energy input from the particle shower, magnet with its superconducting to normal--conducting transition, and the helium that permeates the LHC magnets in the interstitial voids between both cable strands and insulation layers. Models pertaining to the LHC mainly account for the presence of helium in the magnets in one of two ways; 1) they assume no heat transfer directly from magnet to helium, but rather considers helium as an added heat capacity\cite{heliumOnlyHeatCapacity,CLIQ}, or 2) includes surface heat transfer from the superconducting strands into the helium, but assumes no heat transfer within the helium itself\cite{Arjan_CUDI,BreschiGoodCableModellingResults}. The first approach is valid for very fast losses ($\lesssim$10~\us\ range) when it is safe to assume no significant cooling takes place either to the helium nor by way of heat transfer along the cable strands. The second approach is valid for steady state situations where the helium volume is large. At the millisecond time scale of UFOs, when the helium is confined within the LHC magnet cables, however, the validity of the models have not been experimentally investigated to give the assumptions physical basis. The lacking validity of modelling only a single strand with local helium cooling was confirmed by a discovery made after analysing the orbit--bump quench test in the LHC done in 2011\cite{orbitBump_2011}. 

The orbit--bump test revealed that for a purposefully induced loss in the 10~ms time range the model accounting for helium cooling of the magnet cable strands severely underestimates the amount of energy needed to quench. The model predicts that an energy deposition of about 50 to 80~\mJ\ should be sufficient to cause a quench, while the lower limit found during the test is 198~\mJ, for which no quench occurred, and an upper limit of 405~\mJ\ for which a quench was observed\cite[Tab. V]{bernhardsBeamInducedPaper_withFactor4}. 

In order to better understand how helium cooling works on the millisecond time scale in confined volumes, new experimental work is needed. Section \ref{sec:theory} expands on both theory and background from previous experimental work, but in summary, the shortcomings of current understanding are the following; 1) the standard surface heat transfer model was developed for steady state heating into large volumes of helium and we need to give physical basis for using the same heat transfer expression for time--dependent modelling; 2) while prediction of the critical heat flux beyond which helium boiling begins is possible, the film boiling onset, with associated loss of cooling capability, is heavily dependent on local geometrical conditions.

Beyond this we will expand the millisecond--timescale data available in literature, and investigate two minor effects that may change heat transfer in the Kapitza regime; 1) a heat transfer dependence on orientation, and 2) a heat transfer dependence on position along a heater. Furthermore, no reliable Kapitza fit parameters were previously available for stainless steel.

The measurement campaign presented in this paper was conducted in two main steps; 1) gather steady state data from heat transfer to an open bath of He~II in order to validate the setup against expected behaviour; and 2) explore the transient behaviour of surface heat transfer to an open bath of He~II, including the transition to film boiling. 

Validation of the setup in steady state is done by fitting Kapitza parameters to our results between applied heating power densities between 0 and 85~\kW, and comparing these with those found in literature. The transient measurements rely on steps in applied heating power density, up to 85~\kW, to investigate transient heat transfer. We compare our results with the millisecond--timescale data available in literature.
		
	\section{Theory and Background} \label{sec:theory}
		In this paper, we consider heat transfer from a narrow, rectangular heater strip cooled from one side by a large volume of He~II. For analysis of results, two main heat transfer characteristics are necessary;
\begin{itemize}
	\itemsep-1pt
	\item
		Heat transfer from the hot surface to the cold He~II for heat fluxes, or heating durations, that do not trigger the onset of film boiling. We call this the Kapitza regime under both steady state and transient conditions;
	\item
		Heat transfer from the surface to He~II for heat fluxes, or heating durations, that do cause the onset of film boiling. We call this the film boiling onset regime.
\end{itemize}
Van~Sciver provides reviews of these topics, and only the most relevant aspects are discussed herein (see Section 7.5 for the Kapitza regime, and 7.6 for film boiling, in Ref. \cite{VanSciver}).

\subsection{Kapitza Heat Transfer} \label{sec:kapitza}
	Claudet and Seyfert initially proposed the phenomenological expression that describes heat transfer in the Kapitza regime, which will be used herein\cite[Eq. 1]{claudet_seyfert_standardKapitzaExpression};
	\begin{equation} \label{eq:kapitza}
		Q_\text{K} = \aK\ \left( \Ts^{\nK} - \Tb^{\nK} \right),
	\end{equation}
	where \QK\ is the Kapitza heat flux, \Ts\ is the temperature of the heater at heater--He~II interface, referred to as the surface temperature, \Tb\ is the bath temperature of He~II far from the heater, and \aK\ and \nK\ are two fit parameters. This expression fits data for large heat fluxes, on the order of 1~\kW\ and up.
	
	The two fit parameters depend on the heater material as well as the local surface conditions of the heater, and for any given heater, if high accuracy is desired, dedicated measurements must be done to obtain them. Claudet and Seyfert's original measurements showed that copper heaters whose surfaces were prepared identically showed the same variation in measurement results as heaters where the surface preparation was purposefully different (such as after baking or annealing). From across literature there are, however, ranges; \aK\ tends to be in the range 200 to 1300 \aKUnit, and \nK\ in the range 2 to 4\cite{claudet_seyfert_standardKapitzaExpression,GoodlingIrey_possiblePrecursorToClaudetAndSeyfertsExpression,KashaniVanSciver_analyticalExpressionKapitzaModel,ClementFrederking_highHeatFluxExperiment,Shiotsu_92_AuMnWire,Taneda_copperAndSteel}. From the phonon radiation limit, describing the largest possible heat transfer rate across the interface, \nK\ is considered to have an upper limit of 4, stemming from the $T^3$--dependence of the density of phonon states in the Debye approximation\cite[p. 108]{kittel_solidStatePhysics}. A physically consistent theory explaining the Kapitza resistance is the acoustic mismatch theory originally developed by Khalatnikov\cite[Chap. 23]{Khalatnikov_book}. Here, the actual acoustic impedance mismatch between the solid heater material and the He~II is considered, as well as the non--zero reflection probability of a phonon incident upon the interface. This AM theory consistently underestimates measured results, and several attempts have been made to bring it in line with observations. 
	
	Two important avenues attempting to explain the discrepancy are; 1) the presence of an adsorbed/solidified layer of helium atoms on the heater surface that improves acoustic matching\cite{Challis_denseHeliumLayer,opticalPhononMode}, and 2) random surface roughness of the interface that increase the effective heat transfer area\cite{little_kapitzaBetweenSolids,adamenkoAndFuks_roughness,shiren_surfaceRoughness}. Khater and Szeftel merged the two approaches\cite{layerAndRoughness_khaterAndSzeftel} and found convincing agreement with Anderson \etal's old measurements below 1~\K\cite{Anderson_comparisonWithKhater}. Ramiere \etal\ recently found that Adamenko and Fuks's surface roughness model\cite{adamenkoAndFuks_roughness} gives excellent agreement with their measurements on a silicon single--crystal over a wide range of temperatures\cite{ramiere_comparisonWithAdamenkoAndFuks}. The surface roughness considered for these kinds of models is on the order of the phonon wavelength so as to scatter phonons significantly better than the unmodified AM theory. This means roughness amplitudes on the order of a few nanometres at temperatures around 2~\K. This is about two orders of magnitude smaller than the surface roughness expected on steel after high--grade mechanical polishing\cite{steelSurfaceRoughness}.
	
	\vspace{-10pt}
	\subsubsection{Day--to--day Variations} \label{sec:dayToDayVariation}
		\vspace{-5pt}
		An important note about measurements pertaining to heat transfer in He~II is that results have a layer of uncertainty tied to the variation of the Kapitza resistance over long time scales. Rawlings and van~der~Sluijs find in their study on steady and transient heat transfer in He~II, during the early days of large--heat--flux experiments, that they needed to repeat their measurements multiple times over several days until they could obtain results within 5\%\ of each other for the same applied heating power densities\cite{RawlingsAndVanDerSluijs_severalDays}.
		
	\subsubsection{Orientation Dependence}
		In the AM theory of Kapitza heat transfer there is no dependence on gravity, which makes intuitive sense since the mechanism is governed by the transmission of thermal phonons across the interface between a heater and He~II. Phonons are governed by the strong inter--atomic potentials in matter, and are not appreciably affected by a weak gravitational potential\cite[p. 83--84]{kittel_solidStatePhysics}. 
	
\subsection{Film Boiling in He II}
	The Kapitza regime persists until the heat flux across the interface exceeds some critical value, $Q_\text{crit}$, defined as the steady state heat flux above which a film of gaseous helium covers the heater surface. The following relationship governs the heat transfer;
	\begin{equation} \label{eq:filmBoiling}
		Q_\text{FB} = a_\text{FB} \left( \Ts - \Tb \right),
	\end{equation}
	where $Q_\text{FB}$ is the film boiling heat flux, and $a_\text{FB}$ is a coefficient that depends on heater material and configuration (typically 200 to 1000 \unit{\watt\per\square\metre\kelvin}\cite[Table 7.5, ``flat plate'' entries]{VanSciver}). Heat transfer in the film boiling regime tends to be chaotic; even if the heater surface is coated with a continuous layer of vapour, bubbles may form and depart, causing unpredictable local behaviour.
	
	That said, after Kobayashi \etal\ found that the Kapitza regime persists for some time even after the steady state critical heat flux is exceeded\cite{Kobayashi_originalTransientKapitzaObservation}, a large amount of experimental work was done to study the onset of film boiling in He~II for various geometries\cite{VanSciver_transientInLongCoiledPipe,Gradt_transientKapitzaRegime,WangEtal_wireHeater,Shiotsu_modifiedCylindricalPeakHeatFlux,Shiotsu_92_AuMnWire,ShiotsuEtal_transientWire,TatsumotoEtal_originalFlatPlateCriticalHeatFlux,ShiotsuEtal_heaterAtEndOfDuct,Tatsumoto_peakHeatFluxFlatPlate}. Several expressions to determine the critical heat flux were found, and the one Tatsumoto \etal\cite{TatsumotoEtal_originalFlatPlateCriticalHeatFlux} propose for a flat rectangular plate in an open bath is relevant here;
	\begin{equation} \label{eq:peakHeatFlux_plates}
		Q_\text{crit} = K \left[ \frac{2}{\frac{Lw}{2(L + w)}} \int_{\Tb}^{\Tl} \frac{1}{f(T)}\text{d}T \right]^{1/3},
	\end{equation}
	where $K$ is a fit factor equal to 0.58, $L$ is the length of the heater, $w$ its width, and $f^{-1}(T)$ is the thermal conductivity function of turbulent He~II;
	\begin{equation}
		\begin{split}
			f^{-1}(T)	&= g(\Tl) \left[\left(\frac{T}{\Tl}\right)^{6.8} \left(1 - \left(\frac{T}{\Tl}\right)^{6.8}\right) \right]^3,\\
			g(\Tl) 		&= \frac{\rho^2 {s_\lambda}^4 {\Tl}^3}{A_\text{GM}(\Tl)},\\
		\end{split}
	\end{equation}
	with $s_\lambda \simeq 1559$~\unit{\joule\per\kilo\gram\kelvin}, and $A_\text{GM}(\Tl) \simeq 1150$~\unit{\metre\second\per\kilo\gram}. Note, this expression for the thermal conductivity function was modified from the traditional one where the exponent is 5.7, not 6.8, and $A_\text{GM}(\Tl)$ was taken as 1450, not 1150~\unit{\metre\second\per\kilo\gram}\cite[Eq. 7.2]{VanSciver}. Sakurai \etal\ originally proposed this modified version\cite{Sakurai_modifiedCylindricalPeakHeatFlux}.
	
	The typical observation from these measurements is that after applying a heating power density \Qapp, there is a rapid temperature rise of the heater, in accordance with the Kapitza regime (Equation \eqref{eq:kapitza}), which flattens out and remains constant for some time $\tau_\text{Kapitza}$. After this quasi--steady state, the heater temperature shoots up, as film boiling starts. $\tau_\text{Kapitza}$ is called the quasi--steady state Kapitza regime life--time, and it relates to \Qapp\ in one of two main ways;
	\begin{equation} \label{eq:propToQMinFour}
		\begin{split}
			\tau_\text{Kapitza} &\propto \Qapp^{-4} \quad \text{weak heating},\\
			\tau_\text{Kapitza} &\propto \Qapp^{-2} \quad \text{strong heating},\\
		\end{split}
	\end{equation}
	where the delineation between weak and strong heating (though, always above $Q_\text{crit}$), is geometry dependent. The strong--heating behaviour appears to be exclusive to He~II channels\cite{SeyfertEtal_concentricCylindersTimeToBoilingAndTimeDependentSimulation} or heating sufficiently strong that boiling onset starts after less than about 1~\ms\cite{ShiotsuEtal_heaterAtEndOfDuct}. Note also that the time to boiling onset may not follow Equation \eqref{eq:propToQMinFour} until the applied heating power density is as much as a factor 2 above $Q_\text{crit}$.
	
\subsection{Transient Measurements}
	Transient measurements on the millisecond time--scale not intended to investigate second--sound heat transfer have previously only been done as part of the aforementioned studies on transition to film boiling. Generally, the time--dependent data is only presented summarily, such as by Gradt \etal\ (see Ref. \cite{Gradt_transientKapitzaRegime}) and Shiotsu \etal (see Ref. \cite{ShiotsuEtal_transientWire}) with no emphasis on the initial temperature rise, nor on any time--dependent modelling efforts to explain the behaviour. Gradt \etal\ show a time--dependent measurement that appears to take about 0.5 to 0.6~\ms\ to reach the quasi--steady state before then seeing the onset of film boiling after another 1~\ms. Shiotsu \etal's measurements take about 0.4~\ms\ to reach the quasi--steady state. Once film boiling is established, the further temperature rise in both measurement sets appears roughly linear with time, though none of their plots go beyond 10~\ms\ at most.

	\section{Setup}
\begin{figure}[t]
	\centering
	\includegraphics[width=\columnwidth]{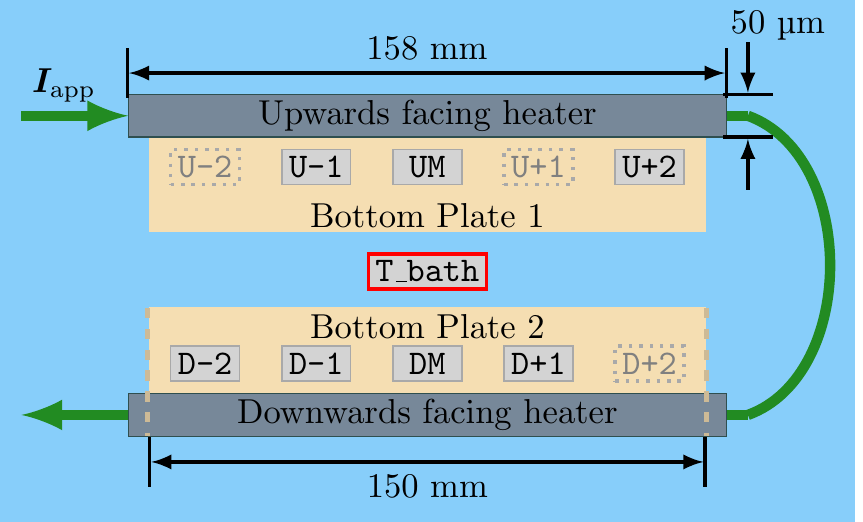}
	\caption{Diagram of the measurement setup. The light blue outer rectangle represents the helium bath. Light grey rectangles represent the \cx\ sensors. \sensor{D}: downwards facing heater sensors. \sensor{U}: upwards facing heater sensors. The number refers to the sensor's relative left (negative) or right (positive) of the middle sensor (\sensor{UM} or \sensor{DM}). The red--edged rectangle represents the reference probe. Edge sensors are 12.5~mm from the edge of their PEEK plates, and 31.25~mm apart. The heater strips are 3~mm wide (in the plane of the paper). Dashed and greyed out sensors broke during assembly or cooldown.}
	\label{fig:diagramOfSamples}
	\vspace{-15pt}
\end{figure}

		Figure \ref{fig:diagramOfSamples} schematically represents the setup placed into the cryostat. The key features in each sample are the stainless steel heater strips, heated by passing current through them, and the \cx\ sensors used to the measure temperatures. Steel was chosen for three main reasons; 1) its high electrical resistance, meaning a relatively low current leads to strong heating; 2) for being easy to work with by hand, relevant for assembly of the setup; and 3) having low thermal conductivity which limits longitudinal heat flow in order to observe potential temperature variation along the heater strips. 

Bare chip \cx\ sensors by Lake Shore Cryotronics were chosen for their excellent thermal response time and temperature sensitivity\cite{lakeshore_cernoxOverviewPaper,Fuzier_fastCernox}. Sensor labelling refers to Upwards or Downwards heater orientation and the indexed position relative to the middle sensor in each plate. The plates themselves are made of glass--fibre filled PEEK, in order to better match the thermal contraction of the heater strips, and thus prevent delamination during cooldown. \sensor{T\_bath}\ represents the pre--calibrated probe used both for in--situ calibration of the other sensors, and as the temperature reference for control of the helium bath temperature.


\begin{figure}[ht]
	\centering
	\includegraphics[width=\columnwidth]{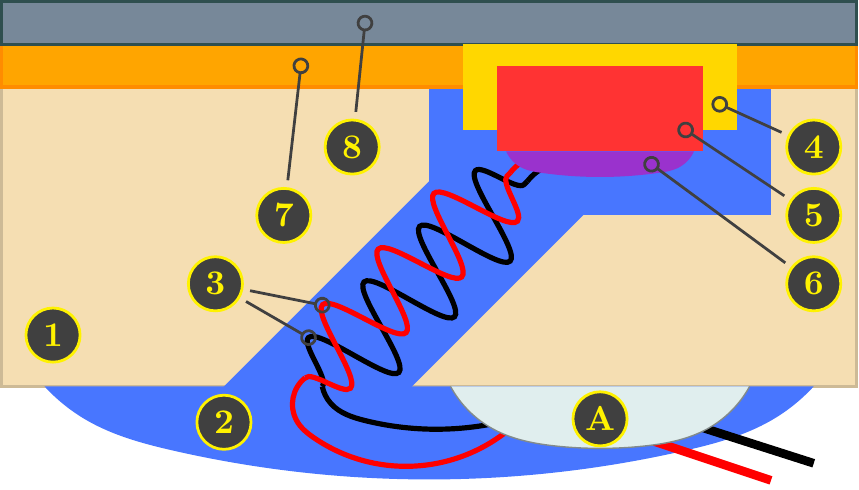}
	\vspace{-10pt}
	\caption{Schematic representation of the region around the \cx\ temperature sensors. Label (1): Glass--fibre filled PEEK. (2) \ecco. (3): copper sensor lead wires. (4): GE 7031 varnish. (5): sapphire sensor substrate. (6): EPO--TEK H20E epoxy. (7): kapton insulation tape. (8): stainless steel heater strip. (A): Soldering point where thin sensor leads join larger sensor lead attachments (one anchor point for each lead wire).}
	\label{fig:regionAroundSensors}
	\vspace{-10pt}
\end{figure}

Figure \ref{fig:regionAroundSensors} shows schematically the immediate surroundings of an individual \cx\ sensor (material parameters are provided in Appendix \ref{app:materials}, and thickness approximations are discussed in Section \ref{sec:x-ray});
\begin{itemize}
	\itemsep-1pt
	\item[\textbf{(1)}]
		PEEK (poly--ether ether ketone), filled with 30\%\ (by volume) glass fibre, oriented so the fibres are parallel to the axis of thermal contraction. The thermal path between the sensors and the back of the sample is dominated by the thin copper leads, so PEEK material parameters are not relevant for the steady state or transient thermal modelling presented in sections \ref{sec:steadyStateResults} and \ref{sec:transientResults};
	\item[\textbf{(2)}]
		\ecco\ epoxy used to fill in holes, chosen for its He~II leak--proofness\cite{whyUseEccobond};
	\item[\textbf{(3)}]
		Copper sensor leads (two per sensor), attached by manufacturer. Diameter 63.5~\um, and length between sensor and thermal anchor (A) 20~mm;
	\item[\textbf{(4)}]
		GE 7031 varnish used to attach \cx\ sensors to the underside of the heater strip. Approximate layer thickness 25~\um. A discussion of this dimension is given in Section \ref{sec:x-ray};
	\item[\textbf{(5)}]
		Sapphire substrate making up the bulk of the \cx\ sensor body. Thickness 203~$\pm$25~\um;
	\item[\textbf{(6)}]
		EPO--TEK H20E silver filled epoxy used by Lake Shore Cryotronics to attach leads to sensor body. Approximate thickness (separating sensor from sensor lead) 20~\um;
	\item[\textbf{(7)}]
		Kapton tape lining the underside of the heater strip, in which a hole is cut so sensors can attach directly to the stainless steel heater. This means kapton, like PEEK (1), is not in the thermal path considered in the modelling. Thickness 100~\um;
	\item[\textbf{(8)}]
		Stainless steel heater strip. Thickness 50 $\pm$2~\um;
	\item[\textbf{(A)}]
		Soldering points (one per sensor lead) joining the sensor leads to larger external lead attachments, acting as thermal anchoring for the \cx\ sensors.
\end{itemize}

The thermal path considered for simulations in sections \ref{sec:steadyStateResults} and \ref{sec:transientResults} runs from the top/surface of the heater strip (8), down through the varnish (4), then the sapphire of the \cx\ sensor (5), then the EPO--TEK (6), and finally the sensor lead wires (3). The wires end in the thermal anchor (A). Note that all interfaces between materials along this thermal path are between a solid and a liquid (that then solidifies upon curing after assembly); the varnish (4) between the steel heater strip (8) and the \cx\ sensor (5) is liquid during assembly, as is the EPO--TEK (6) and the solder (A). For this reason, we will not consider thermal contact resistances as relevant for the heat transfer modelling as there are no interfaces whose contact depend on solids pressed together.

\subsection{Calibration}
	The \cx\ sensors are calibrated in--situ against the reference probe. The double--bath cryostat used for experiments does not permit temperature control above 4.2~\K, so in this region, only quasi--steady temperature readings from the slow cooldown process are available for calibration purposes. A possible impact of this is a temperature offset caused by the thermal diffusion time between the helium--cooled surface of the heater and the sensitive part of the \cx\ sensors. However, in the temperature range 4 to 50~\K, adding the thermal diffusion times through the various layers of materials gives an overall diffusion time (through 50~\um\ of stainless steel, 25~\um\ of varnish, and 200~\um\ of sapphire) $\tau_\text{diff} \simeq$ 0.1~\ms\ at around 4~\K, growing to 1~\ms\ at around 50~\K. This is much shorter than the time--rate of temperature change of the helium bath during cooldown, which is on the order of 1 to 10~\unit{\milli\kelvin\per\minute}.
	
	The calibration is done by associating the measured resistance of each individual \cx\ sensor at several temperatures (known from the reference probe), and using a cubic spline function to represent this calibration for analysis. This means our calibrated \cx\ sensors have a contribution to their total measurement uncertainty equal to the calibration uncertainty of the reference probe (provided by Lake Shore Cryotronics).
	
	While there is no temperature offset, there is a source of uncertainty due to the transient nature of the calibration data. Only below \Tl\ does the cryostat temperature control permit long--term temperature stability. So each calibration data point collected above \Tl\ is based on data that is changing, albeit slowly, in time. To estimate this additional contribution, we take the calibration spline and apply it to the raw data of each individual point. This gives time--dependent temperature curves for each \cx\ sensor that deviate slightly from the time--dependent reference probe reading. As an example; we have a measurement that gives the calibration point we use at around 3.35~\K. During the measurement the reference probe reading drops from around 3.37 down to 3.23~\K\ over the course of an hour. Outside the narrow time--window we use for the calibration point itself, we find the root--mean--square deviation between the reference and all \cx\ sensors, and then take the uncertainty contribution within this temperature range as the average RMS deviation across all the sensors. The total estimated measurement uncertainty is listed for temperature ranges in Table \ref{tab:uncertainty}. Below 4~\K, the main contribution is from the reference probe uncertainty, while above 4~\K, the main contribution is from the transience of the calibration measurements.
	
	\begin{table}[ht]
		\vspace{-5pt}
		\renewcommand{\arraystretch}{1}
		\caption{Estimated measurement uncertainty $\Delta T$.}
		\label{tab:uncertainty}
		\vspace{-6pt}
		\centering
		\begin{tabular}{r c l r}
			\multicolumn{3}{c}{Range, [\K]} & $\pm\Delta T$, [\mK]	\\ \hline
			1.8		&	---		& \Tl		& 5 \\
			\Tl		&	---		& 4			& 7 \\
			4		&	---		& 6			& 15 \\
			6		&	---		& 50		& 50
		\end{tabular}
		\vspace{-20pt}
	\end{table}
	
\subsection{Instrumentation}
	The data acquisition is split across two systems; one system for triggering current pulses in the heater strips and measuring the resulting voltage across them, and another system to measure the voltage across the temperature sensors, which then are converted to temperature by the calibration spline function. Both systems are run through LabView\textsuperscript{\textregistered}.
	
	For the heater strips, voltage is measured across only the upwards facing heater, and the voltage for the downwards facing heater is obtained by scaling with the resistance ratio between the two strips. To find the resistance ratio, we fed a steady 1~\unit{\ampere} current and measured the voltage across both strips, and then across just the upwards facing heater. This yielded $R_\text{up} = 0.465 \pm 0.001$~\unit{\ohm} and $R_\text{down} = 0.458 \pm 0.001$~\unit{\ohm}. During transient measurements, the strip current is found using these resistances and Ohm's law. Since stainless steel has only a negligible resistivity change with temperature below liquid nitrogen temperatures, this method introduces no uncertainty beyond that of $R_\text{up}$ and $R_\text{down}$. The data acquisition frequency is typically 500~\unit{\kilo\hertz}.
	
	For the \cx\ sensors, a four--lead circuit is used, where two sensors share a single 10~\uA\ excitation current source but have individual voltage measurement leads. Sensors \sensor{U-2} and \sensor{U-1} are excited by the same source. The same goes for the pairs \sensor{UM} and \sensor{U+2}, \sensor{D-2} and \sensor{D-1}, and \sensor{DM} and \sensor{D+1}. Typical data acquisition frequency is 40~\unit{\kilo\hertz} per sensor. In order to provide a smooth output current, the excitation current sources each have a low--pass filtering capacitor measured to 0.929~$\pm$0.03~\uF\ in parallel with the output terminals. The effect of this capacitor will be discussed in Section \ref{sec:capacitanceCompensation}.
	
	\begin{figure*}[t]
		\centering
		\begin{subfigure}[t]{0.48\textwidth}
			\includegraphics[width=\columnwidth]{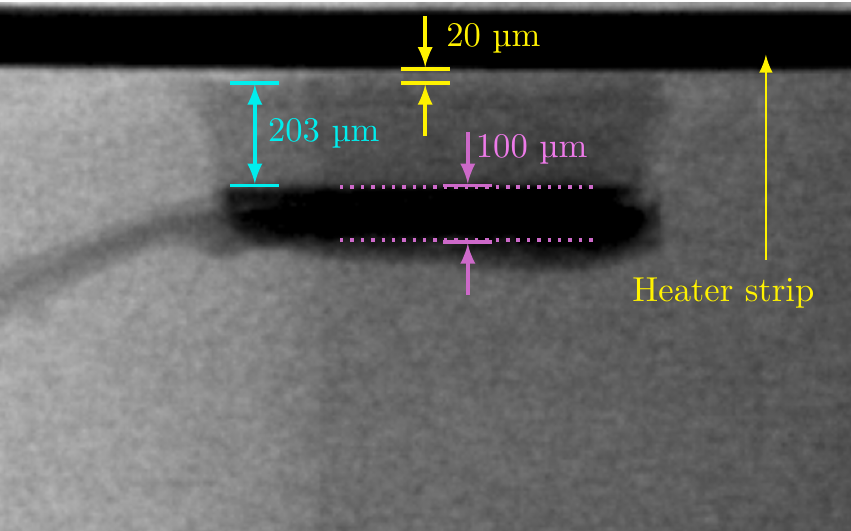}
			\vspace{-8pt}
			\caption{X--ray image of \sensor{U-1}. Distance between heater and sensor base is about 20~\um. Purple dimensions indicate the thickness of the EPO--TEK beads.}
			\label{fig:x-ray_of_U-1}
		\end{subfigure}
		\hfill
		\begin{subfigure}[t]{0.48\textwidth}
			\includegraphics[width=\columnwidth]{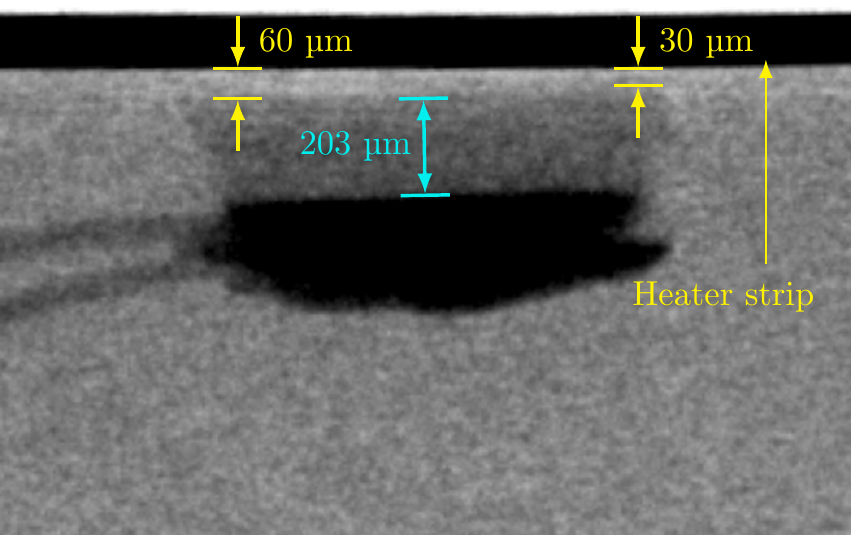}
			\vspace{-8pt}
			\caption{X--ray image of \sensor{UM}. Distance between heater and sensor base is goes from about 60 to 30~\um, for an average of 45~\um.}
		\end{subfigure}
		\vspace{-8pt}
		\captionsetup{width=\textwidth}
		\caption{X--ray images of \sensor{U-1} and \sensor{UM}, showing approximate distances. The imaging resolution was 6~\um. Cyan dimensions are used as the scale in the images.}
		\label{fig:xRays}
		\vspace{-5pt}
	\end{figure*}
	
\subsection{Region Around Sensors} \label{sec:x-ray}
	To better characterise the geometry around the sensors in the real samples after assembly and curing, we have some x--ray images taken at 6~\um\ resolution. Figure \ref{fig:xRays} shows the images for \sensor{U-1} and \sensor{UM} with approximate dimensions indicated. Regions of denser material show up with darker pixels in the images, but note that since the EPO--TEK is loaded with silver particles, the x--rays see an effectively denser material than that expected from the average epoxy density.
	
	The known height of the \cx\ sensors, 203~\um, is used as the scale in both images, and this way we can estimate the distance between the heater strip and the sensors. \sensor{U-1} is about 20~\um\ from the heater, while \sensor{UM} is a little tilted, going from 60 to 30~\um\ away. We have similar images for \sensor{U-2} and \sensor{U+1}. \sensor{U-2} is flush with the heater, to within the image resolution; there cannot be zero varnish, so we take 6~\um\ as the varnish length here. \sensor{U+1} is, like \sensor{UM}, tilted, going from 50 to 0~\um; again we take 6~\um\ in place of zero.
	
	The final varnish length we use for analysis is the root--mean--square value of estimated dimensions, without first taking the average of the estimates for titled sensors. This leads to $l_\text{varnish}$ = 35$\pm$6~\um. The uncertainty is taken as the image resolution.

	Also in Figure \ref{fig:x-ray_of_U-1} is the approximate thickness of the EPO--TEK lead attachment bead, shown as $\simeq$100~\um, which means, assuming the copper leads are in the middle of the bead, there is 20~\um\ of EPO--TEK between the sensor sapphire bulk and the sensor leads. The variation in EPO--TEK thickness will be considered between 10 and 30~\um\ when estimating uncertainty and parameter sensitivity in thermal modelling.
	
	The copper lead wires come attached from Lake Shore Cryotronics with length around 25~mm, and after assembly, where part of the wire end is used for soldering to the anchor, 20~mm of wire runs between the thermal anchor and the EPO--TEK attachment points on the sensors. There is, however, a variation in the length of each lead. The x--ray images were taken with much larger field of view than shown in Figure \ref{fig:xRays}, and from these we estimate the leads vary between 18 and 22~mm long.
	
	In summary, the one--dimensional thermal path considered as the reference domain for modelling purposes consists of 50~\um\ of stainless steel, 35~\um\ of varnish, 200~\um\ of sapphire, 20~\um\ of EPO--TEK, and 20~mm of copper. Appendix \ref{app:heatEquation} describes the implementation of the one--dimensional heat equation we use. 
	
	There are two copper leads per sensor, of diameter 63.5~\um, giving a cross sectional ratio to the sensor cross section of $2 \cdot A_\text{lead} / A_\text{sensor}$, with $A_\text{lead}$ = $\pi (63.5\,\um / 2)^2$ and $A_\text{sensor} = 762 \times 965$~\unit{\square\micro\metre}. This ratio is used to lower the effective thermal conductivity of the copper leads. We also account for the sensor leads passing through \ecco, which poses an additional heat capacity.
	
	
\subsection{Measurement Procedure}
	All measurements follow the same general approach; trigger a step in current, from zero to some value, through the heater strips, and measure the voltages across the heater strip and the \cx\ sensors. All our measurements are thus transient, and steady state results are extracted from measurements after all voltages/temperatures have stabilised. Once current is turned off again, we let all temperatures settle at the bath temperature before applying a new step.
	
	To represent the heating power we use an equivalent applied heating power density, \Qapp;
	\begin{equation} \label{eq:Qapp}
		\Qapp = \frac{{V_\text{strip}}^2}{R_\text{strip}} \frac{1}{A_\text{strip}},
	\end{equation}
	with $R_\text{strip}$ = 0.465~\unit{\ohm} for the upwards facing heater, and 0.458~\unit{\ohm} for the downwards facing heater, $A_\text{strip}$ = $158 \times 3$~\unit{\square\milli\metre}\ the heat transfer area of a strip. $V_\text{strip}$ is simply the measured voltage $V_\text{meas}$ across the upwards facing heater, while for the downwards facing heater the measurement must be scaled by the resistance ratio between the two strips; $V_\text{strip,\,down}$ = $V_\text{meas} \cdot (R_\text{down}/R_\text{up})$.
	
	This calculated applied heating power density is not the heat flux crossing the heater--to--helium interface. \Qapp\ represents the applied volumetric heating power expressed as watts per square metre. The instantaneous or steady state heat flux across the interface depends on the heat loss to the back of the sample as well as the heat capacities of materials. The steady state analysis in Section \ref{sec:steadyStateResults} finds the heat loss through the steady state heat equation and the measured sensor temperatures, while the transient analysis in Section \ref{sec:transientResults} finds the interface heat flux using the applied heating power density and the time--dependent heat equation.
	
	Measurements were done at two bath temperatures; at 1.9~\K\ we go up to \Qapp\ = 85~\kW, while at 2.05~\K\ we go up to \Qapp\ = 68~\kW.
		
\subsection{Capacitance Compensation} \label{sec:capacitanceCompensation}
	The presence of the filtering capacitors in parallel with the current output from the \cx\ sensor excitation current sources mean we must account for this effect during the initial stages of the temperature transient. The \cx\ sensors in each excitation circuit constitute total electrical resistance on the order of 20 to 30~\kOhm\ for an initial temperature of 1.9~\K. The filtering capacitors are 0.929~\uF. During the rapid heating of the sensors, their resistance falls to much lower values, meaning the electrical circuit approaches that of a charged capacitor releasing energy into a resistor. The expected electrical time--constant is on the order of 20 to 30~\ms, while the thermal time--constant we intend to measure is on the order of 1~\ms.
	
	During the transient time between turning on the applied heating power and reaching a steady state, the excitation current in the \cx\ sensors is composed of two independent parts;
	\begin{equation} \label{eq:i_sensors}
		i_\text{sensors}(t) = I_\text{EX} + C\df{V_\text{C}(t)}{t},
	\end{equation}
	where $I_\text{EX}$ is the steady 10~\uA\ current, fed by the ideal part of the excitation circuit, and $V_\text{C}(t)$ is the time--dependent voltage across the filtering capacitor $C$. This voltage is the sum of voltages across the two sensors in the same excitation circuit.
	
\begin{figure}[ht]
		\centering
		\includegraphics[width=\columnwidth]{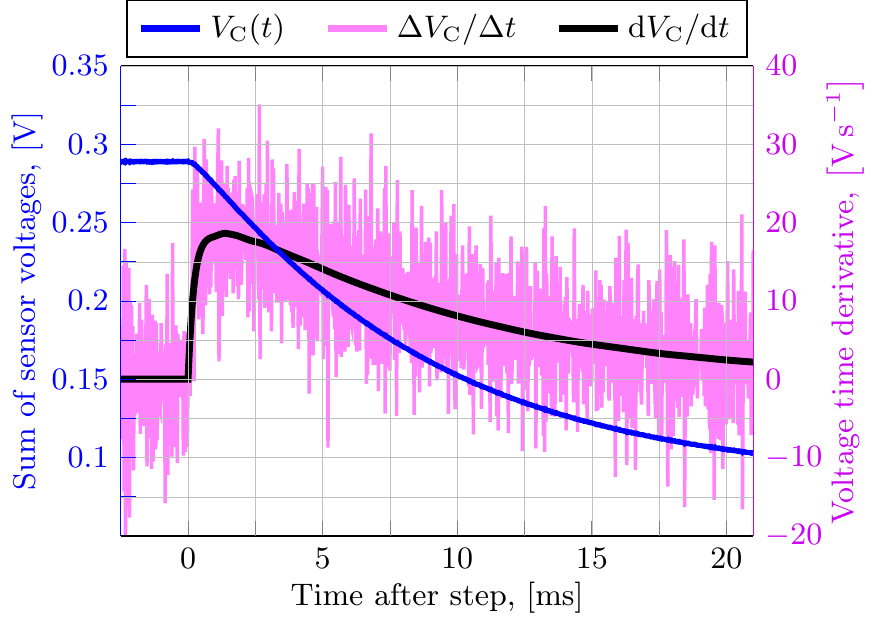}
		\vspace{-18pt}
		\caption{Sensor voltages and voltage time derivative to be used in Equation \eqref{eq:i_sensors}. Left axis: sum of voltages across \sensor{UM} and \sensor{U+2}. Right axis, raw and smooth voltage time derivative.}
		\label{fig:voltageDerivative}
		\vspace{-10pt}
	\end{figure}

	The left axis of Figure \ref{fig:voltageDerivative} shows the sum, \legendEntry{$V_\text{C}(t)$}, of the raw voltage signals of sensors \sensor{UM} and \sensor{U+2} from a measurement during a step in applied heating power density. The right axis shows the numerical time derivative \legendEntry{$\Delta V_\text{C}/\Delta t$} of this voltage sum. The shape of the voltage time derivative is representative of all measurements made; only the signal amplitude varies with applied heating power density. As is expected, the numerical differentiation introduces significant noise, so some form of filtering is necessary. Our approach is roughly split in three, relying on the use of the Savitzky--Golay filter from the Python function \verb|scipy.signal.savgol_filter|, and the low--pass Butterworth filter from \verb|scipy.signal.butter|;
	\begin{enumerate}
		\item
			Using the \verb|savgol_filter|, we directly obtain the numerical derivative from the voltage signal, instead of using the very noisy $\Delta V/\Delta t$ approach;
		\item
			The low--pass \verb|butter| filter is used to obtain two filtered versions of the \verb|savgol_filter| result; one with a high cut--off frequency, that captures the early region around the peak of the response, and one with a low cut--off frequency that captures the long tail of the signal;
		\item
			The \verb|savgol_filter|, due to it representing a numerical derivative, tends to give non--zero voltage derivative values at the time of the step. \emph{Before} the step in heating power, the sensors are at their initial temperature, and no voltage change should be present. Therefore, a polynomial is fitted to the first millisecond of the \verb|savgol_filter| result such that all values before the step are zero.
	\end{enumerate}
	These three parts combine, with smooth transitions from one region to the next, to give the much improved voltage time derivative curve \legendEntry{$\text{d}V_\text{C}/\text{d}t$}. Each of the four sensor excitation circuits will have their own voltage time derivative curve. The same behaviour is seen also when power is turned off at the end of a step, although the signal is much weaker.
	
	The smooth $\text{d}V_\text{C}/\text{d}t$ is used to find the time--dependent sensor excitation current from Equation \eqref{eq:i_sensors}, and this is used to calculate the instantaneous sensor resistance of each individual sensor (since we measure the individual sensor voltages). This compensated resistance is converted to temperature using the calibration spline fits for each sensor.

	\section{Steady State Results} \label{sec:steadyStateResults}
\begin{figure}[t]
	\centering
	\includegraphics[width=\columnwidth]{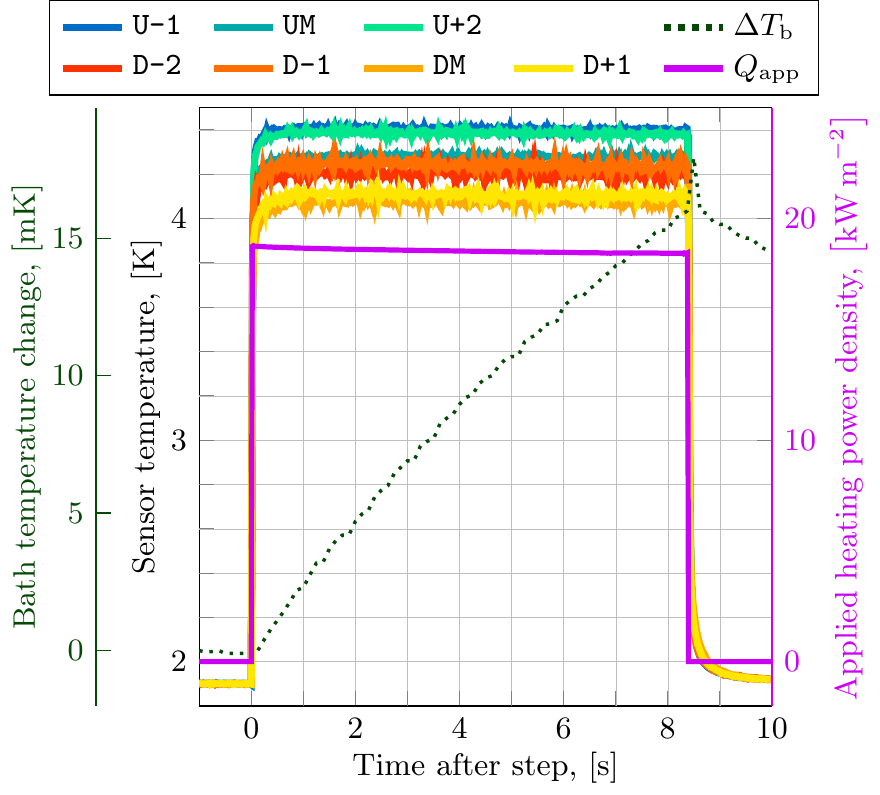}
	\vspace{-18pt}
	\caption{Representative measurement for a single step in applied heating power density, whence steady state data is extracted. Initial bath temperature is 1.9~\K. For clarity, only one in every thousand data points is shown.}
	\label{fig:representativeStepMeasurement}
	\vspace{-12pt}
\end{figure}

		Figure \ref{fig:representativeStepMeasurement} shows a representative measurement result where a step up to 4.4~\Amp\ is applied at time $t=0$, and turned off after about 8 seconds. For plot clarity, only the upwards facing heater has its \Qapp\ shown.

There is a slight bath temperature increase, peaking around 16~\mK, because the two heater strips supply a large amount of heat over the course of the test. For the highest applied heating power densities used, around 85~\kW, the peak bath temperature rise is about 50~\mK.

Steady state data is taken as an average over the last 2 to 3 seconds. The bath temperature considered for steady state analysis is the average in the same time--window.

\subsection{Steady State Heater Surface Temperature}
	The temperatures in Figure \ref{fig:representativeStepMeasurement} are those from the \cx\ sensors themselves, which, recall, are separated from the surface by about 200~\um\ of sapphire, 35~\um\ of varnish, and 50~\um\ of stainless steel. So, as with any measurement where the heater surface is not directly instrumented (or itself used as a sensing element), the heater surface temperature must be obtained from the measurements by way of a thermal model. We use the steady state heat equation with temperature dependent thermal conductivities and the Python library \verb|lmfit| to determine the surface temperature at each sensor location. We know the thermal anchor at the back of the sample is at the bath temperature, and we know the applied heating power density. The \verb|lmfit| routine then guesses the value of the surface temperature under the condition of minimising the difference between the measured sensor temperature and the temperature at the location of the sensor in the simulated domain.
	
	\subsubsection{Our Day--to--Day Variation}	
\begin{figure}[t]
			\centering
			\includegraphics[width=\columnwidth]{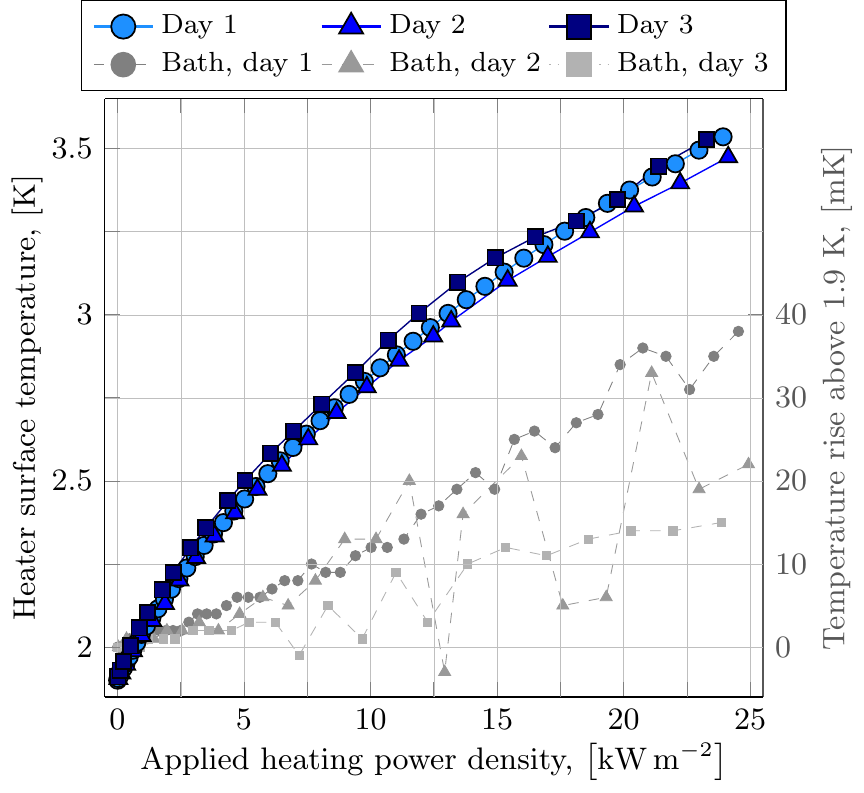}
			\vspace{-18pt}
			\caption{Heater surface temperatures of sensor \sensor{UM}, for applied heating power densities up to 25~\kW, for three separate measurement days. Initial bath temperature is 1.9~\K.}
			\label{fig:steadyState_threeDays}
			\vspace{-12pt}
		\end{figure}

		Figure \ref{fig:steadyState_threeDays} shows the modelled surface temperature for \sensor{UM} as it varies with applied heating power density up to around 25~\kW, for three separate measurement days. The main measurement campaign, with the highest heating powers, was \legendEntry{Day 3} (dataset is truncated here). Between measurement days, several days passed, and going from \legendEntry{Day 1} to \legendEntry{Day 2}, the setup spent a weekend at 4.2~\K. It is not clear what causes the small variation across long time scales, but, as mentioned in Section \ref{sec:dayToDayVariation}, this is not unexpected. The estimated measurement uncertainty from Table \ref{tab:uncertainty} is smaller than the observed variation by about a factor 4 across all measurements. Between \legendEntry{Day 1} and \legendEntry{Day 2}, the calibration was redone after finding an apparent calibration shift of around 10~\mK, but as seen, the day--to--day variation is larger than this.
		
\subsection{Kapitza Model Fit}
	\begin{figure}[t]
		\centering
		\includegraphics[width=\columnwidth]{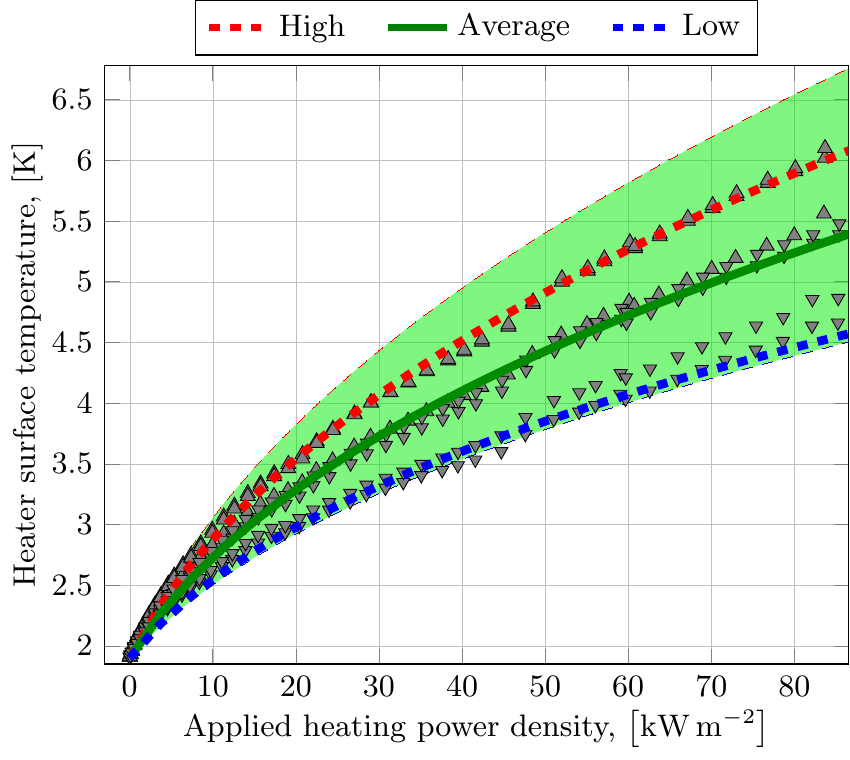}
		\vspace{-18pt}
		\caption{Range of heater surface temperatures within the estimated parameter space, represented by fits to Equation \eqref{eq:kapitza}. Fit parameters are shown in Table \ref{tab:kapitzaFitParameters}.}
		\label{fig:kapitzaWithErrors}
		\vspace{-10pt}
	\end{figure}

	Figure \ref{fig:kapitzaWithErrors} shows calculated surface temperatures from the full data set from \legendEntry{Day 3}, for all sensors, up to an applied heating power density of about 85~\kW. Upwards--pointing triangles belong to sensors on the upwards facing heater. The temperature variation between sensors at the highest applied heating is 1.46~\K, from 4.67~\K\ on \sensor{DM} to 6.13~\K\ on \sensor{U+2}, which is roughly in line with the variation of 1.4~\K\ (from 4.78 to 6.16~\K) Claudet and Seyfert found for their various copper samples at \Qapp\ = 80~\kW\ (Ref. \cite{claudet_seyfert_standardKapitzaExpression}). Kashani and Van~Sciver also found variations between identical samples (around 0.17~\K\ at 50~\kW), though not nearly as large as Claudet and Seyfert (Ref. \cite{KashaniVanSciver_analyticalExpressionKapitzaModel}). There does appear to be an orientation dependence present, since the downwards facing heater shows consistently lower temperatures. No such effect is expected from the theory of Kapitza conductance, however. The variation between sensors could also stem from there being significant differences in the Kapitza conductance from one heater surface location to another. An important caveat to this is that there is a certain parameter sensitivity to the method used to obtain the heater surface temperatures. This sensitivity is explored in the following section.
	
	The three curves in Figure \ref{fig:kapitzaWithErrors} represent fits to the Kapitza heat transfer expression in Equation \eqref{eq:kapitza}. The fits are made using, again, the Python package \verb|lmfit|, with \aK\ and \nK\ as free parameters. The input to the fitting procedure is the set of simulated heater surface temperatures belonging to one sensor at a time, together with the corresponding surface--to--helium heat fluxes adjusted for the heat leak backwards to the bath. The heat leaks are estimated from the heater surface temperature calculation, and represents between 2 and 3\%\ of the total applied heating power density. The \verb|lmfit| routine then varies \aK\ and \nK\ looking for the combination of parameters that minimise, in a least--squares sense, the difference between the simulated surface temperature and that calculated from Equation \eqref{eq:kapitza}.
	
	Table \ref{tab:kapitzaFitParameters} gives the Kapitza parameters for the three curves. \legendEntry{High} is the fit to \sensor{U+2}, which shows the highest temperatures, while \legendEntry{Low} is the fit to \sensor{DM}, which shows the lowest temperatures. \legendEntry{Average} is the the fit to the average of all seven sensors at each heating power density. The green shading represents the area between the limits in Equation \eqref{eq:kapitzaParameters}.
	\begin{equation} \label{eq:kapitzaParameters}
		\aK\ =\ 1316.8 \pm 10\%,\ \ \nK\ =\ 2.528 \pm 10\%
	\end{equation}
	Equation \eqref{eq:kapitzaParameters} represent the first reliable Kapitza parameters published for stainless steel. Note that the upper limit in the plot goes with the lower values in the range in Equation \eqref{eq:kapitzaParameters}.
	
	\begin{table}[ht]
		\centering
		\caption{Least--squares fit parameters for \sensor{U+2} and \sensor{DM}, and the average temperature at each heating power density, to Equation \eqref{eq:kapitza}.}
		\label{tab:kapitzaFitParameters}
		\vspace{-10pt}
		\begin{tabular}{r c c}
			Curve & \aK $\left[\aKUnit\right]$ & \nK \\ \hline
			High & 1335.5 & 2.35 \\
			\textbf{Average} & \textbf{1316.8} & \textbf{2.53} \\
			Low & 1213.6 & 2.86 
		\end{tabular}
	\end{table}

	\subsubsection{Parameter Sensitivity} \label{sec:sensitivity}
		Along and between the heater strips, looking now only at the highest applied heating power density in Figure \ref{fig:kapitzaWithErrors} for the sake of clarity, there appears to be considerable temperature variation, on the order of 0.5 to 0.6~\K\ on the same heater, and 1~\K\ between the upwards and downwards facing heaters. 
		
		The method used to obtain the heater surface temperatures has a certain sensitivity to variations in the input parameters. The most important parameters that may impact the calculated heater surface temperature are; 1) the length of the copper leads (Label 3 in Figure \ref{fig:regionAroundSensors}) between sensor and anchor (Label A); 2) the thickness of the EPO--TEK layer (Label 6) between the sensor and the sensor leads; 3) the thickness of the varnish layer between the heater strip and the sensor (Label 4); and 4) the thermal conductivity of stainless steel (Label 8).
		
		
		Comparisons will be made between the reference temperatures at the highest applied heating power density in Figure \ref{fig:kapitzaWithErrors} and the temperature found for the same heating power after changing one of the parameters.
		\paragraph{Copper lead wires}
			Using a lead length of 22~mm lowers the surface temperature about 95~\mK, or 1.9\%, below the reference, while using a length of 18~mm increases the surface temperatures by about 110~\mK, or 2.2\%\, above the reference.
		\paragraph{EPO--TEK silver filled epoxy}
			Using an EPO--TEK length of 10~\um\ gives surface temperatures 70~\mK, or 1.4\%, above reference, while using a length of 30~\um\ gives temperatures 65~\mK, or 1.2\%, lower.
		\paragraph{GE 7031 varnish}
			Using a varnish layer thickness of 29~\um\ gives surface temperatures about 240~\mK, or 4.2\%, lower than reference, while using 41~\um\ gives temperatures 230~\mK, or 4.2\%, higher than reference.
		\paragraph{Stainless steel}
			Typical thermal conductivity measurement uncertainty is on the order of $\pm$5\%~\cite[p.~13]{lorenzNumber}; by lowering the stainless steel conductivity by 5\%\ we find heater surface temperatures 460~\mK, or 8.0\%, lower than the reference, while increasing the conductivity by 5\%\ increases the surface temperatures by 350~\mK, or 6.1\%. We assume the thermal conductivity is uniform along and between heater strips, so the effect of this uncertainty cannot help explain the sensor variations in Figure \ref{fig:kapitzaWithErrors}. This uncertainty simply moves all curves up or down.
	
	\subsubsection{Surface Temperature Variation}
		The total variation in local dimensions around the sensors lead to about $\pm$400~\mK\ uncertainty in the calculated heater surface temperatures. This is a smaller variation in temperature than that seen in Figure \ref{fig:kapitzaWithErrors}, meaning that the spread in surface temperatures cannot fully be explained by the uncertainty range we have characterised. 
		
		Looking only at the temperature variation between sensors on the same heater strip, $\pm$400~\mK\ is large enough to give the same calculated heater surface temperatures, so we cannot conclude there is significant temperature variation along the same sample. 
		
		Claudet and Seyfert's results on identical copper heaters show variation between samples similar to what we see between upwards and downwards facing heaters. We cannot, therefore, conclude there is an orientation dependence of the Kapitza conductance.

	\section{Transient Results} \label{sec:transientResults}
\begin{figure}[ht]
	\centering
		\includegraphics[width=\columnwidth]{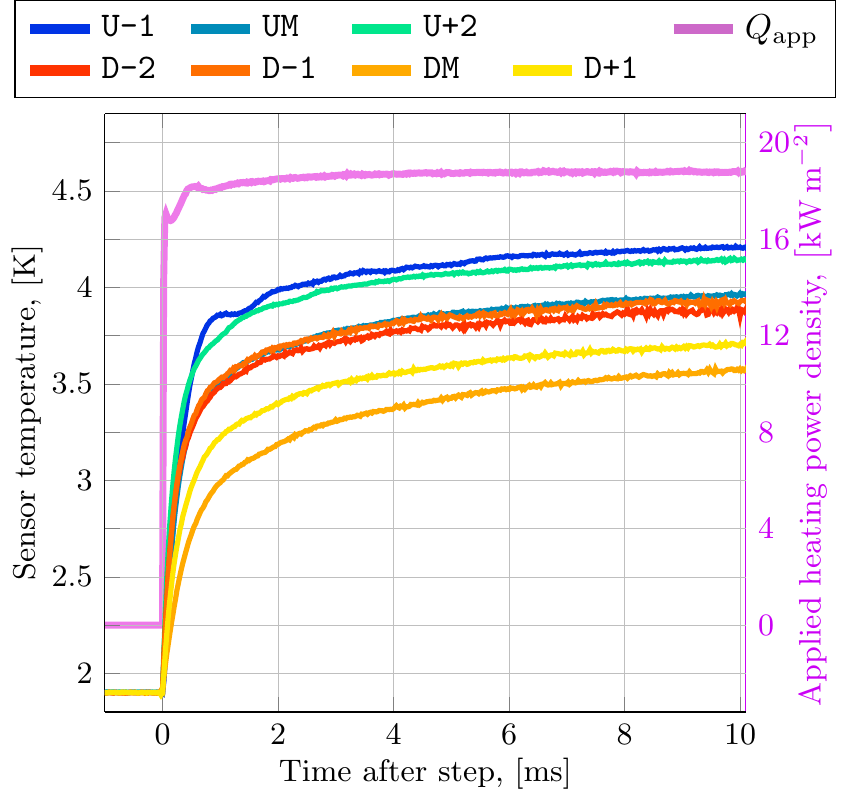}
	\vspace{-18pt}
	\caption{Representative transient measurement. Same test as in Figure \ref{fig:representativeStepMeasurement}, after compensating for the filtering capacitor in the current excitation sources, showing only the first 10~\ms.}
	\label{fig:representativeTransientMeasurement_noSim}
	\vspace{-15pt}
\end{figure}

		Figure \ref{fig:representativeTransientMeasurement_noSim} shows the same test as that in Figure \ref{fig:representativeStepMeasurement}, focusing on the first 10~\ms, after applying the capacitive compensation method described in Section \ref{sec:capacitanceCompensation}. Note that these are sensor temperatures. 

The first important observation is that all curves look very similar, the main difference being the steady state temperature they approach. This similarity represents an important validation that there are no small leaks or reservoirs of He~II influencing the measurements; if there were, we would see the characteristic impact of the large helium heat capacity at the lambda transition. Furthermore, the impact would be different between sensors, since such reservoirs or leaks would not be of equal size and location for each sensor.

For all heating power densities tested, the temperature rise follows this general behaviour; rapid initial rise lasting about 1~\ms, then a long, slow rise towards the steady state value. It takes on the order of a full second to reach the final steady state. 

The tests in a bath of 2.05~\K\ where the applied heating power density is sufficiently high to see film boiling onset has a different behaviour once boiling starts to develop, but show the same characteristic early temperature rise (discussed more in Section \ref{sec:boiling}).

A general remark about all measurements shown in this section; unless otherwise noted, during the time windows shown in figures and considered for analysis, the bath temperature remained constant at its initial value.

\subsection{Thermal Time--Constants}	
\begin{figure}[ht]
		\centering
			\includegraphics[width=\columnwidth]{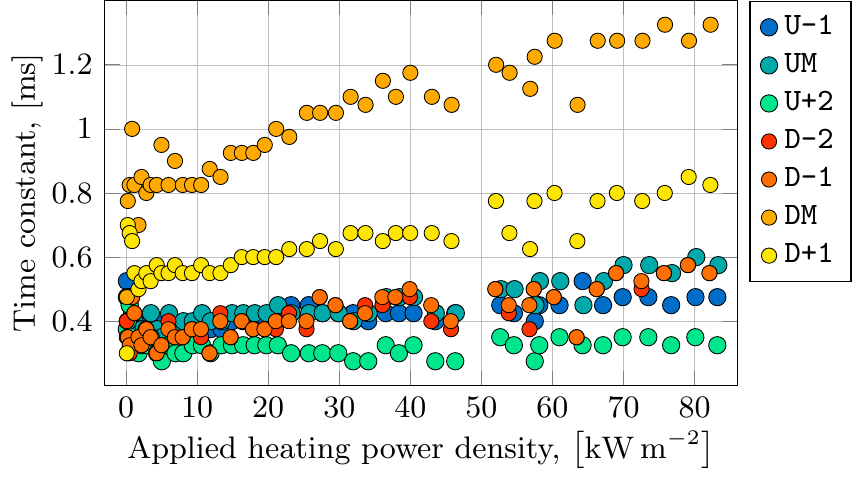}
		\vspace{-18pt}
		\caption{Estimated thermal time constants for all steps in applied heating power density in a bath of 1.9~\K.}
		\label{fig:timeConstants1p9K}
		\vspace{-10pt}
	\end{figure}

	To quantify the initial temperature rise we define a thermal time constant for the curves, $\tau$, as the time it takes each sensor, for each heating power, to reach the temperature $(1 - 1/e)\left[T(t=10\,\ms) - \Tb\right] + \Tb$. Looking at \sensor{DM} in Figure \ref{fig:representativeTransientMeasurement_noSim}, this is the time at which the temperature has reached about 2.97~\K, which is around 0.9~\ms. Figure \ref{fig:timeConstants1p9K} shows this time constant for all sensors and all applied heating power densities from tests in a bath of 1.9~\K. Sensors \sensor{DM} and \sensor{D+1} are significantly slower than the others, and also slow further for growing \Qapp. Otherwise, the sensors all show time constants between 0.3 and 0.5~\ms. For tests in a bath of 2.05~\K, there is no significant difference in the thermal time constant as compared with Figure \ref{fig:timeConstants1p9K}. The largest heat capacity in the system is that of stainless steel, which only grows by about 10\%\ from 1.9 to 2.05~\K. This percentage increase in the thermal time constant would only be on the order of 30 to 50~\us, which is roughly the same as the sampling period of 25~\us, and thus too small a difference to reliably measure. The increase we see for higher heating power densities stems from the temperature going up, meaning the heat capacity goes up, slowing the temperature rise. 
	
	Note that thermal time constants on the order of 0.3 to 0.5~\us\ is quite in line with those found by Gradt \etal\ (see Ref. \cite{Gradt_transientKapitzaRegime}) and Shiotsu \etal (see Ref. \cite{ShiotsuEtal_transientWire}).
	
	So, an obvious question is why \sensor{DM} and \sensor{D+1} deviate so much from the other sensors; thermal time constants are dominated by the total heat capacity of materials in the thermal path. For \sensor{DM} and \sensor{D+1} the various dimensions discussed in Section \ref{sec:sensitivity} must be towards their upper limits, slowing down heat transfer from increased length and thermal mass.
	
\subsection{Simulating a Step}
	\begin{figure}[t]
		\centering
		\includegraphics[width=\columnwidth]{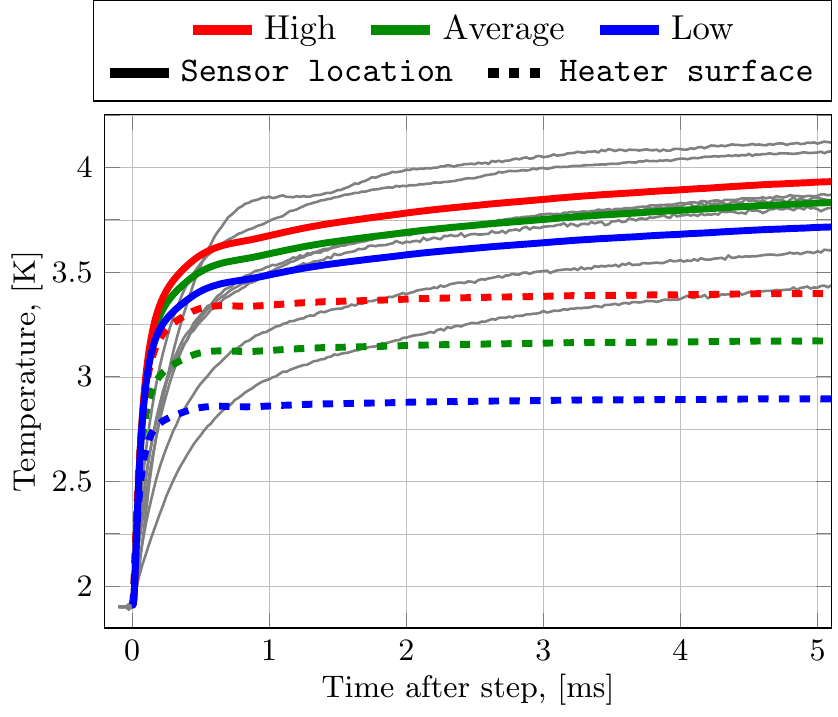}
		\vspace{-18pt}
		\caption{Representative transient measurement with simulated temperatures. Same test as that shown in Figure \ref{fig:representativeTransientMeasurement_noSim}. Grey curves are measured sensor temperatures. The solid coloured curves are the temperatures at the sensor location in the transient simulation, while the dashed coloured curves are the heater surface temperatures in the simulation. \legendEntry{High}, \legendEntry{Average}, and \legendEntry{Low} refer to the Kapitza parameters in Table \ref{tab:kapitzaFitParameters}.}
		\label{fig:representativeTransientMeasurement_withSim}
		\vspace{-10pt}
	\end{figure}
	
	Using the time--dependent heat equation (see Appendix \ref{app:heatEquation}), with Equation \eqref{eq:kapitza} defining the instantaneous heat flux across the interface between the heater surface and the bath of He~II, we can simulate how the thermal response of the setup using the measured heater strip voltage as a volumetric heat source in the stainless steel heater. This simulation does not consider any dynamics in the helium, beyond using the measured bath temperature as input to the Kapitza heat transfer expression. Figure \ref{fig:representativeTransientMeasurement_withSim} shows the result of this simulation for the three Kapitza parameter sets in Table \ref{tab:kapitzaFitParameters}, compared with the measurement shown in Figure \ref{fig:representativeTransientMeasurement_noSim}. We use standard material lengths; 50~\um\ of steel, 35~\um\ of varnish, 200~\um\ of sapphire, 20~\um\ of EPO--TEK, and 20~mm of copper leads. The applied heating power density, not shown in the plot for clarity, is that from the upwards facing heater; the downwards facing heater has lower resistance, and therefore, about 1.5\%\ lower heating power density than that used in the simulation. The grey curves in the figure are the measured sensor temperatures shown in Figure \ref{fig:representativeTransientMeasurement_noSim}.
	
	The rapid early temperature rise, which is faster in simulations than measurements, stems from the Kapitza heat transfer mechanism needing a substantial temperature difference across the heater--to--helium interface to move appreciable amounts of heat. Before this temperature difference is established, the heater strip warms up nearly adiabatically. After the initial rise, after between 1 and 1.5~\ms, simulations are in excellent agreement with the measured temperatures. The slow rise after 1.5~\ms\ is dominated by the thermal mass behind the sensor needing time to heat up. Note that the simulated heater surface temperature remains nearly steady even if the sensor temperature keeps growing.
	
\begin{figure}[ht]
		\centering
		\includegraphics[width=\columnwidth]{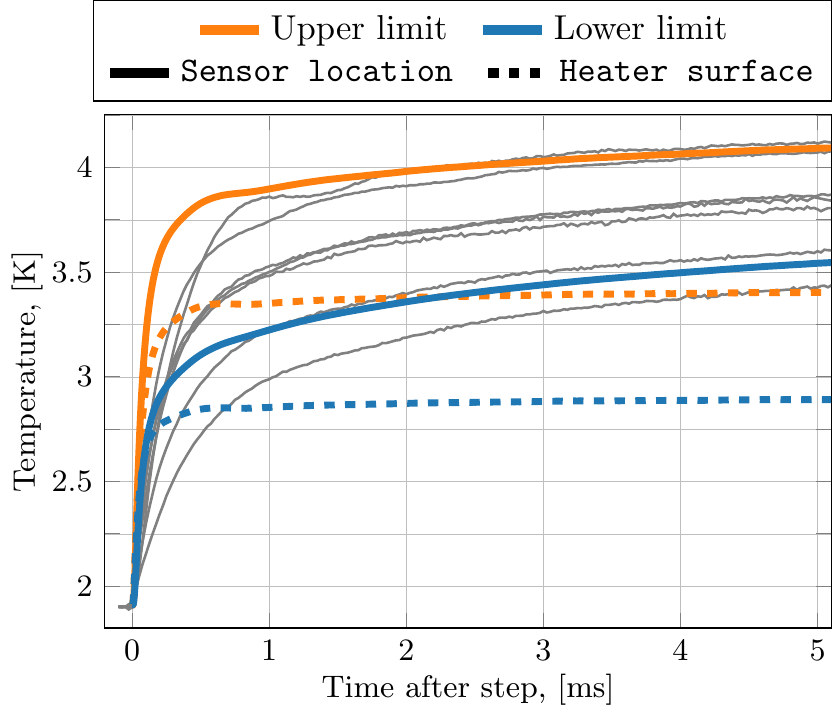}
		\vspace{-18pt}
		\caption{Like Figure \ref{fig:representativeTransientMeasurement_withSim}, with simulations using the extrema in the parameter space. \legendEntry{Upper limit} is the simulation that gives the highest simulated sensor temperature within the parameter space, and \legendEntry{Lower limit} the lowest.}
		\label{fig:representativeTransientMeasurement_withSim_extrema}
		\vspace{-10pt}
	\end{figure}

	Figure \ref{fig:representativeTransientMeasurement_withSim_extrema} shows simulation results using parameters that give the largest sensor temperature variation within the estimated parameter ranges. \legendEntry{Upper limit} uses \legendEntry{High} Kapitza parameters from Table \ref{tab:kapitzaFitParameters} with 29~\um\ varnish, 30~\um\ EPO--TEK, and 22~mm copper lead length. \legendEntry{Lower limit} is on the other end of the spectrum; \legendEntry{Low} Kapitza parameters, 41~\um\ varnish, 10~\um\ EPO--TEK, and 18~mm copper lead length. 
	
	The simulated heater surface temperatures are only marginally different from those shown in Figure \ref{fig:representativeTransientMeasurement_withSim} because the Kapitza parameters used are the same; the small heat leak backwards does not represent a sufficiently large fraction of the total heat flow to lower the surface temperature appreciably. The simulated sensor temperatures match the range of measured sensor temperatures very well, but, again, only after about 1 to 1.5~\ms. 
	
	\subsubsection{Slowing Down the Simulated Temperature Rise}
		We seek an explanation for why superconducting magnets subject to losses in the millisecond time--scale need much larger energy depositions to quench than what is actually observed. We see from figures \ref{fig:representativeTransientMeasurement_withSim} and \ref{fig:representativeTransientMeasurement_withSim_extrema} that simulations, using Equation \eqref{eq:kapitza} to represent the transient Kapitza cooling mechanism, consistently show faster temperature rises than the measured values. The discrepancy we see between simulation and measurement clearly means the system heats up slower than expected. The question is if the discrepancy is due to a feature of the setup not correctly accounted for in the model, or because the Kapitza expression represents a less effective cooling than what is really going on during the first millisecond of the transient. 
		
		
		A simple test of this is to replace the Kapitza boundary condition with a fixed temperature, representing a form of perfect cooling. With the reference simulation parameters, but the heater surface temperature clamped to the bath temperature, we get an initial sensor temperature rise essentially identical to the other simulated sensor temperature curves in figures \ref{fig:representativeTransientMeasurement_withSim} and \ref{fig:representativeTransientMeasurement_withSim_extrema}. The difference being that with the heater surface temperature fixed at 1.9~\K, the sensor temperature approaches a lower steady state value than that seen in measurements. That the ``perfect cooling'' gives a similar initial temperature rise is as expected; the early temperature rise is effectively adiabatic as the thermal gradient within the steel develops in order to move the required heat flux to balance the applied heating power density.
		
		This result, where excessive cooling still does not slow down the temperature rise, points towards there being effects unaccounted for when translating the real three--dimensional setup to the simplified one--dimensional model. The discrepancy exists at all tested applied heating power densities. For the highest heating power densities (70~\kW\ and up), the discrepancy appears to last a little longer; as long as 2~\ms\ for 85~\kW. This points towards the model not adequately accounting for the effective heat capacity of the region around the sensor; heat capacity depends strongly on temperature, growing by a factor 5 to 10 just going from 2 to 4~\K. Stronger heating leads to higher temperatures reached more quickly, and therefore higher heat capacities for the materials involved.

\subsection{Film Boiling Onset} \label{sec:boiling}
\begin{figure}[ht]
		\centering
			\includegraphics[width=\columnwidth]{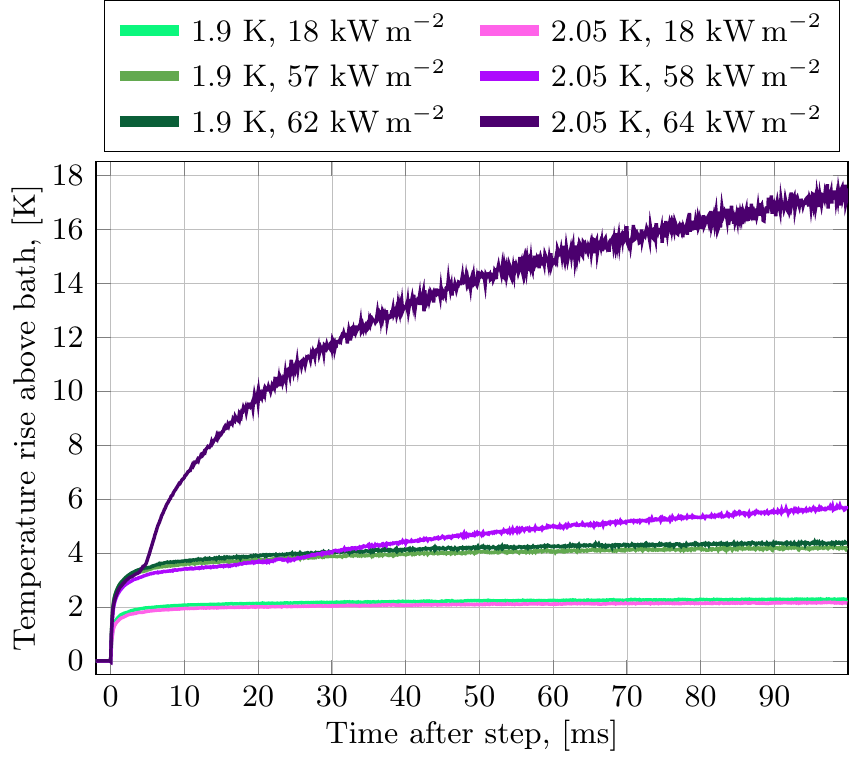}
		\vspace{-18pt}
		\caption{Measured sensor temperatures for the same applied heating power densities in baths of 1.9 and 2.05~\K, represented by Sensor \sensor{UM}. The variation in \Qapp\ stems from the current source not supplying exactly the same current from one test to another.}
		\label{fig:compare1p9And2p05KAroundBoiling}
		\vspace{-10pt}
	\end{figure}

	In a bath of 2.05~\K\ we see the onset of film boiling for applied heating power densities above 58~\kW. Figure \ref{fig:compare1p9And2p05KAroundBoiling} shows the sensor temperature rise above the initial bath temperature of Sensor \sensor{UM} for the same three heating power densities in 1.9 and 2.05~\K\ baths. The small discrepancy in \Qapp\ stems from slight variations in the output current from the power source between tests. 
	
	Two immediate observations are clear; 1) at low applied heating power density, the transient behaviour is essentially indistinguishable between the two bath temperatures, which is as expected; in the Kapitza expression, the bath temperature changes the resulting heat flux only very little for high heater surface temperatures. And 2) near the critical heat flux, the film boiling onset is a very gradual process; during the test that gave \legendEntry{2.05~K, 58~\kW}, the sensor temperature reaches about 24~\K, undoubtedly a fully developed film boiling situation, and yet the transition seen in the figure is very smooth, as opposed to that for the higher applied heating power density.
	
	Applying Equation \eqref{eq:peakHeatFlux_plates} to our setup, with $L$ = 158~mm and $w$ = 3~mm, gives an estimated critical heat flux, at which boiling should start, of 47~\kW. The lowest applied heating power density for which we see boiling is \Qapp\ = 55~\kW, which starts after around 300~ms. During these 300~\ms, the bath temperature rose by 2.5~\mK. The difference between the critical heat flux from Equation \eqref{eq:peakHeatFlux_plates} and the one we observe can easily be explained by a slight change in Tatsumoto \etal's fit parameter $K$; changing its value to 0.66, instead of 0.58, gives a critical heat flux of 54~\kW.
	
	Shiotsu \etal\ find that at about 70~\kW\ the time to film boiling onset on a flat--plate--heater in an open bath is on the order of 5~\ms\ in a bath of 2~\K\ and 0.8~\ms\ in a bath of 2.1~\K\ (see Ref. \cite{ShiotsuEtal_heaterAtEndOfDuct}). Our measurements in a bath of 2.05~\K\ find film boiling onset after between 3.1 and 4.5~\ms\ for \Qapp\ around 68~\kW, entirely in line with Shiotsu \etal.
	
\begin{figure}[ht]
		\centering
			\includegraphics[width=\columnwidth]{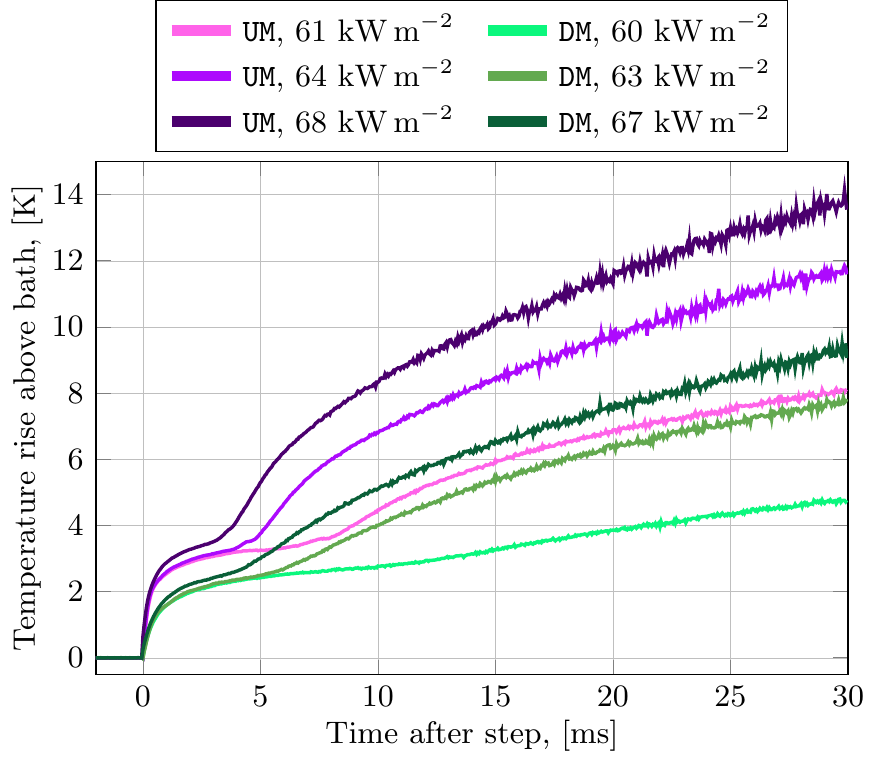}
		\vspace{-18pt}
		\caption{Measured sensor temperatures for the same heating power densities, represented by both middle sensors (\sensor{UM} and \sensor{DM}). The variation in \Qapp\ stems from the small difference in heater strip resistance.}
		\label{fig:compareUP_and_DOWNAroundBoiling}
		\vspace{-10pt}
	\end{figure}

	Figure \ref{fig:compareUP_and_DOWNAroundBoiling} shows how, in a bath of 2.05~K, the upwards and downwards facing heaters differ. The Kapitza quasi--steady life--time is clearly visible; \legendEntry{\texttt{UM}, 61~\kW}, for instance, flattens out between 2.5 and 7.5~\ms, before then showing the characteristic rise due to film boiling onset. There also appears to be a slight difference in how long it takes for boiling to start; for an upwards facing plate, a bubble forming at the surface will have the help of gravity to detach and carry off energy, while for a bubble forming on the surface of a downwards facing heater, the bubble must both fight gravity as its centre of mass is moved downwards in the fluid, and also move sideways, rather than straight up, in order to move away from the heater. This helps explain why boiling starts a little sooner on the upwards facing plate seeing as there is the additional energy barrier of buoyancy to overcome on the downwards facing heater. Now, as more and more bubbles form, it will be harder for an individual bubble on the surface of the downwards facing heater to move sideways without coalescing with another bubble. This means there will be a less defined delineation between a situation with individual bubbles and one where the bubbles form a continuous film as compared with the upwards facing heater where the film only fully forms once bubbles arise across the entire surface at once. These two effects are subtle, as the time to boiling onset becomes more similar for higher heat flux, and the upwards/downwards difference becomes less pronounced. Shiotsu \etal\ show transitions to film boiling from a thin wire that are about equally smooth to ours for the downwards facing heater, while the transition for our upwards facing heaters appear sharper than theirs (see Ref. \cite{ShiotsuEtal_transientWire}). Our measurements transition into a linear temperature rise with time, as theirs do.
	
	During our tests, we typically only let the film boiling regime develop for about 2 seconds. A steady state is not reached within this time, and sensor temperatures increase linearly for most of the test. As we turn off power, the sensors have reached as much as 40 to 45~\K. Temperatures return to the initial bath temperature in a smooth process, lasting as long as 3 to 4 times longer than the duration of the step. The linear temperature rise lasts from about 100~\ms\ after the step until power is turned off. Note that while the sensor temperatures are growing rapidly, the bath temperature hardly changes at all; the peak bath temperature measured during the strongest step is only 20~\mK\ above the initial bath temperature.
	
	The long transition period between onset of film boiling and this linear region (seen in full for \legendEntry{2.05~\K, 64~\kW} in Figure \ref{fig:compare1p9And2p05KAroundBoiling}) is completely smooth, without any obvious regime changes. So, the identifiable heat transfer regimes appear to be the quasi--steady Kapitza regime, before film boiling onset, and the single film--boiling--like regime without any other clear transitions, where the film boiling onset is not a distinct regime

	\subsubsection{Time to Film Boiling Onset}
\begin{figure}[ht]
			\centering
				\includegraphics[width=\columnwidth]{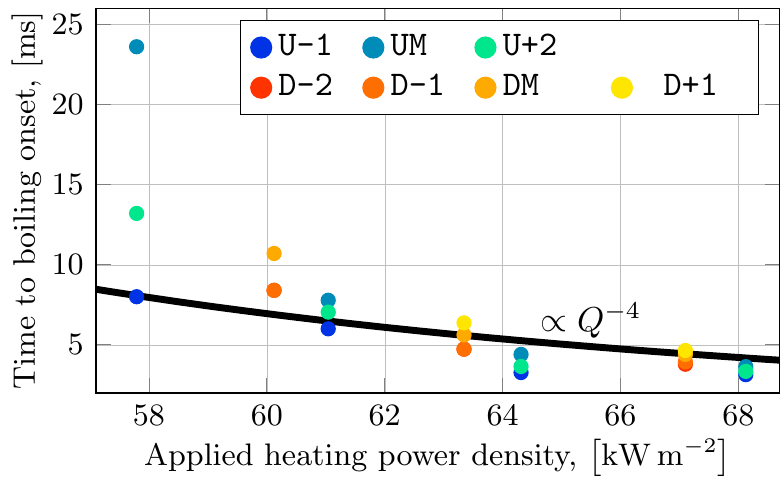}
			\caption{Time to boiling onset as function of \Qapp, showing also a curve $\propto Q^{-4}$.}
			\label{fig:timeToBoil}
		\end{figure}

		Although our setup was not designed to measure the life--time of the Kapitza regime accurately, we can get rough estimates of Kapitza regime life--time, $\tau_\text{Kapitza}$, from the curves in Figure \ref{fig:compareUP_and_DOWNAroundBoiling} by taking $\tau_\text{Kapitza}$ as the time of the kink upwards as the criterion for identifying the film boiling onset. Figure \ref{fig:timeToBoil} shows the film boiling onset time (or Kapitza regime life--time) for all measurements where we saw film boiling onset within a few tens of milliseconds after the step in power. The curve in the figure is an example of a curve $\tau_\text{Kapitza} \propto Q^{-4}$, which our measurements tend to follow as \Qapp\ goes above 60~\kW, like expected from Equation \eqref{eq:propToQMinFour}.
		
	\subsubsection{Simulate Quasi--Steady Life--Time}
\begin{figure}[ht]
			\centering
			\includegraphics[width=\columnwidth]{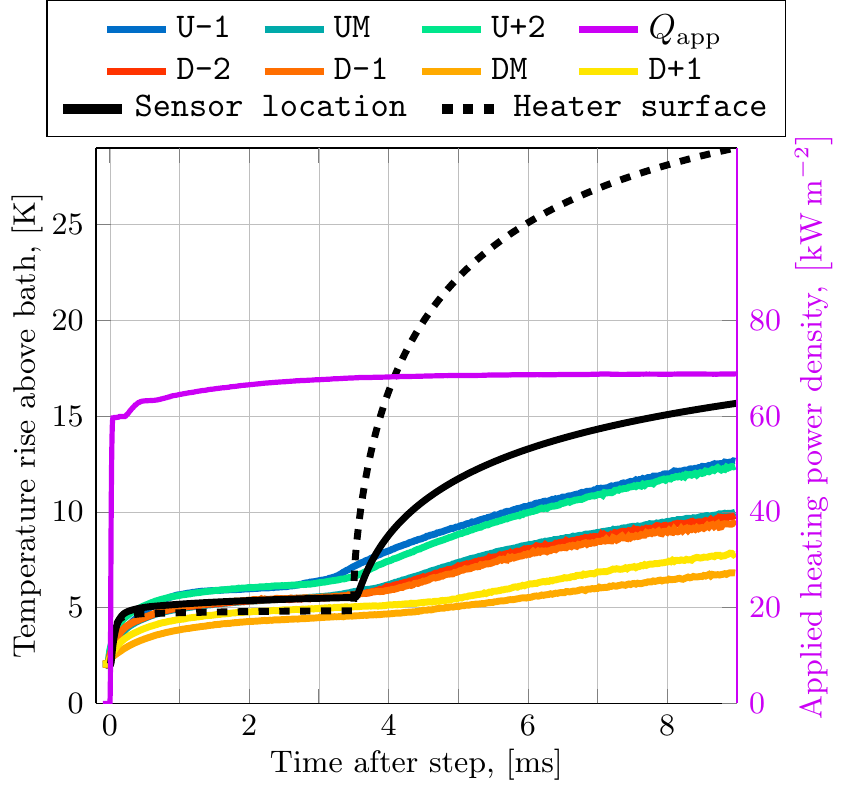}
			\vspace{-18pt}
			\caption{Measured sensor temperatures during step in applied heating power density to 68~\kW, in bath of 2.05~\K, together with simulation using reference parameters.}
			\label{fig:simulateUpToBoiling}
			\vspace{-12pt}
		\end{figure}

		Figure \ref{fig:simulateUpToBoiling} shows the step in applied heating power density up to 68~\kW, in a bath of 2.05~\K\, together with a simulation using reference parameters. At $t$ = 3.5~\ms\ we instantaneously change the heat transfer regime from the Kapitza expression in Equation \eqref{eq:kapitza} to the much weaker film boiling expression in Equation \eqref{eq:filmBoiling} with $a_\text{FB}$ = 500~\unit{\watt\per\square\metre\kelvin}. The quasi--steady Kapitza life--time is obvious; the simulated sensor temperature stabilises between $t$ = 1 and 3~\ms, at which point the film boiling onset is reached. At 3.5~\ms\ we trigger the fully developed film boiling heat transfer regime. An important insight from this very simple modelling approach is that the drop in heat transfer capability when going into film boiling is so large that most of the heat is now transferred backwards through the material stack rather than by the film boiling heat flux itself. It is also clear that the real film boiling onset is a much smoother process than a hard transition from a high to a low heat transfer regime, seeing as even the sharper \sensor{U-1} and \sensor{U+2} sensors see a mellower transition into the film boiling onset temperature rise.
		
		Note also that, save for the discrepancy during the first millisecond, also a step that leads to film boiling can be accurately simulated using the simple steady state Kapitza expression as the cooling boundary condition before film boiling onset.
		
	\section{Conclusion}
		To begin investigating the observation that significantly more heat input is necessary to quench an LHC superconducting magnet than what models have predicted, we built an experimental setup with the aim of making millisecond time--scale measurements of transient cooling of a heater cooled by an open bath of He~II. We confirm the setup behaves as expected in steady state, where the Kapitza heat transfer model is known to work, and quantify the measurement uncertainty from a wide range of parameters. The result is a set of Kapitza fit parameters valid for applied heating power densities \Qapp\ between 1 and 85~\kW; \aK\ = 1316.8$\pm10\%$, and \nK\ = 2.528$\pm10\%$.

Further validation of the experimental setup is found by the critical heat flux being in line with approximate expressions from literature relevant to our heating geometry, and then by seeing that the time from turning on heating power to onset of film boiling follows the expected $\propto Q^{-4}$ behaviour.

The setup allowed us to investigate heat transfer variations along heaters and between upwards and downwards facing heaters. We do not find evidence of significant differences attributable to local surface conditions along the heaters, nor differences attributable to an orientation dependence. The difference between the upwards and downwards facing heater temperatures is in line with what can be expected simply from the two heaters being unique.

The rise time of the initial measured temperature response after a step in applied heating power density is on the order of 0.3 to 0.5~\ms, which is similar to the data found in literature, and we provide considerably more time--resolved data than that which has previously been published.

We have made a time--dependent one--dimensional model representation of the setup. The model's thermal response to a step in \Qapp\ is faster than what we measure during the first millisecond after the step, with an initial rise time about half that of measurements, but after this, the agreement between measurement and model is excellent. This means we confirm the assumption that the steady state Kapitza heat transfer expression can be used for fast transient modelling.
		
	\appendix
		\renewcommand{\thesection}{\Alph{section}}
		\numberwithin{equation}{section}
		\numberwithin{figure}{section}
		\renewcommand\thefigure{\thesection.\arabic{figure}} 
		\section{Material Parameters} \label{app:materials}
\begin{figure*}[ht]
	\centering
	\includegraphics[width=\linewidth]{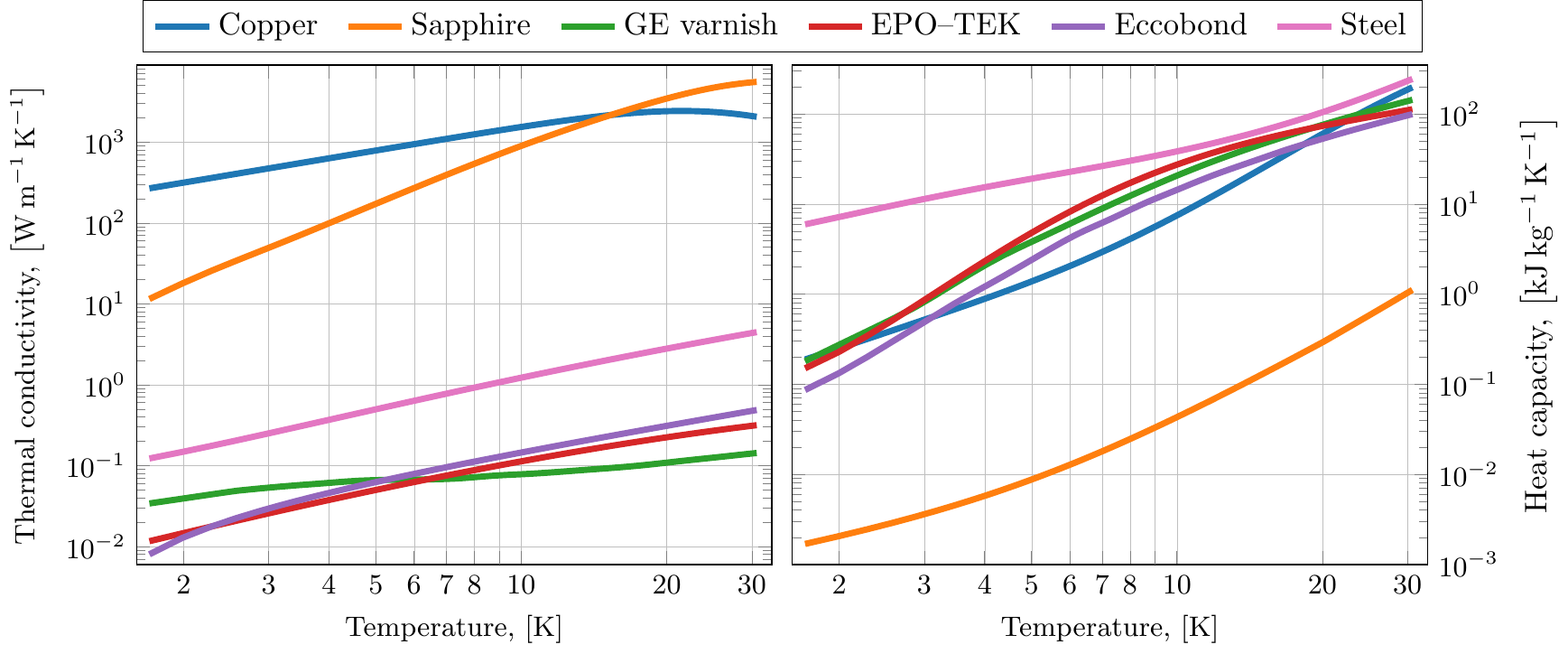}
	\captionsetup{width=\textwidth}
	\vspace{-8pt}
	\caption{Thermal conductivity (left) and heat capacity (right) as used in analysis herein. Plots highlight the most relevant region below 30~\K, though the parameters are known till 100~\K.}
	\label{fig:materialParameters}
	\vspace{-5pt}
\end{figure*}

			Figure \ref{fig:materialParameters} shows the thermal conductivity and heat capacity of the seven materials included in the analysis of data. The figures stop at 30~\K, since measured temperature never go above this, but all materials have known parameters in the range 1.7 to 100~\K.

The data, with densities from Table \ref{tab:density}, comes from the following sources;
\paragraph{Copper}
	Heat capacity is a fit by NIST\cite{NIST_copper}. Note that although the range is from 4 to 300~\K, the expression they propose fit their source data by Stewart and Johnson down to 1~\K\cite[p. 4.112--1]{NIST_Copper_background_C}. Thermal conductivity is a fit function proposed by Simon \etal\cite{NIST_Copper_background_k}, where we assume here that the RRR of copper is 100.

\paragraph{Sapphire}
	Heat capacity is Viswanathan's polynomial fit below 20.0661~\K\cite{Cp_sapphire_upTo20K}, and Fugate and Swenson's polynomial fit above\cite{Cp_sapphire_above20K}. Lake Shore Cryotronics provide thermal conductivity data for sapphire\cite[Figure 1.]{k_Sapphire_LakeShore}, and the curve seen in Figure \ref{fig:materialParameters} is a cubic spline fit to the Lake Shore Cryotronics curve.
	
\paragraph{GE 7031 varnish}
	Heat capacity data for the GE 7031 varnish is taken from a measurement by Heessels\cite{Cp_Varnish}. No fit function is provided, so a cubic spline fit to the data is used. For thermal conductivity, the measurements by McTaggart and Slack are used\cite{k_varnish}. They do not provide fit functions, so cubic splines are fitted to the data.

\paragraph{EPO--TEK H20E}
	This proprietary compound does not have readily available data across the entire relevant temperature range. Heat capacity data up to 9~\K\ is taken from Weyhe \emph{et al.}\cite{silverFilledEpoxy_C_lowTemperature}. Extending the range is done by using heat capacity data for Stycast 2850FT, measured by Swenson\cite[Figure 3]{silverFilledEpoxy_C_highTemperature}. The final heat capacity curve is made by making a fifth degree polynomial fit to the logarithm of Weyhe \etal's data up to 8~\K\ and the Swenson data from 30~\K\ up. Thermal conductivity is a third degree polynomial fit of the logarithm of Amils \etal's data\cite{silverFilledEpoxy_k}.
	
\paragraph{\ecco}
	Neither heat capacity nor thermal conductivity data is available for \ecco\ 286 A/B. The heat capacity used for \ecco\ is that of Stycast 2850FT based on the full temperature range measured by Swenson\cite[figures 4 and 3]{silverFilledEpoxy_C_highTemperature}. A cubic spline fit to this data is used. The thermal conductivity used for \ecco\ is that of \ecco\ 285, a single--component epoxy from the same manufacturer, for which a short measurement set exists between 4 and 8~\K, by Rondeaux \etal\cite{k_Eccobond_285}. This data indicates a linear temperature dependence of the thermal conductivity, and this relationship is extrapolated from 1.7~\K\ up. Since heat capacity is that of Stycast 2850FT, the density of \ecco\ is taken as that of Stycast. This probably leads to a higher thermal diffusion time than what is true, for \ecco's actual density, 1400~\density, is only 60\%\ that of Stycast.
	

\paragraph{Stainless steel}
	The heat capacity of 304 stainless steel used for the heater strips is found in Du Chatenier \etal\ for temperatures below 90~\K\cite{Cp_stainlessSteel_to90K}, and in NIST reference data above 90~\K\cite{NIST_steel_Cp}. The logarithm of Du Chatenier \etal's data is fitted by a fifth degree polynomial, while NIST provide their own fit. 
	
	Thermal conductivity is more complicated. Between 1~\K\ and 1.7~\K, Stutius and Dillinger made measurements on 304 stainless steel\cite{CryogenicSteel_Stutius}, and they quantify the lattice contribution to thermal conductivity. Between 6~\K\ and 110~\K, Hust and Sparks give Lorenz ratio measurements for a compositionally similar steel they call HS(347)\cite[p. II-34]{lorenzNumber}. The Wiedemann--Franz law relate the electrical resistivity to thermal conductivity through the Lorentz ratio\cite[p. 153]{kittel_solidStatePhysics}. The sum of these two constrictions is compared with dedicated measurements on our steel samples around 4.2~\K, in order to scale the Wiedemann--Franz result. The measurements, and thus the thermal conductivity used herein, are a factor 0.9383 lower than the WF result.

\begin{table}[ht]
	\renewcommand\arraystretch{1.2}
	\centering
	\caption{Volumetric density of relevant materials}
	\begin{tabular}{l r}
		Material & Density [\unit{\kilo\gram\per\cubic\metre}] \\ \hline
		Copper\cite{copper_ρ} & 8960 \\
		Sapphire\cite{sapphire_ρ} & 3980 \\
		GE Varnish\cite{GE_varnish_ρ} & 887 \\[5pt]
		EPO--TEK H20E\cite{silverFilledEpoxy_ρ_epoTek} & 2550 \\
		Eccobond/Stycast\cite{silverFilledEpoxy_ρ_stycast} & 2292 \\
		Steel\cite{steel_ρ} & 7955 \\
	\end{tabular}
	\label{tab:density}
\end{table}
		
		\section{Heat Equation} \label{app:heatEquation}
			During a transient where some voltage is measured across the heater strip, the one--dimensional heat equation takes the form,
\begin{equation}\label{eq:heatEq_withSource}
	C_\text{p}(x, T)\rho(x, T) \partialt{T} = \frac{\partial}{\partial x}\left\{ k(x, T) \partialx{T} \right\} + \frac{V_\text{meas}^2}{R_\text{s}v_\text{s}},
\end{equation}
where $R_\text{s}$ is the electrical resistance of a heater strip,  and $v_\text{s}$ is the volume of it.

The two boundary conditions are, 1) a Dirichlet condition at the extreme end of the material stack where \ecco\ touches helium, with temperature fixed to the bath temperature, and 2) a Neumann condition at the cooled surface of the stainless steel heater strip;
\begin{align}
	T(x=x_\mathrm{end}, t) &= T_\text{bath}(t), \label{eq:Diriclet}\\
	\left. \partialx{T} \right|_{x=0} &= \frac{Q_\text{Cooling}(t)}{k_\text{steel}(T(x=0, t))}, \label{eq:Neumann}
\end{align}
where $Q_\text{Cooling}$ is cooling--regime dependent. In the Kapitza regime, for instance, Equation \ref{eq:kapitza} is used. The Neumann condition is implemented as a central difference in the numerical scheme to preserve second order accuracy in space.

To solve Eq. \ref{eq:heatEq_withSource}, a Crank--Nicolson scheme is used. To account for the temperature dependent thermal conductivities, as well as there being interfaces between several materials, the thermal conductivity is evaluated between adjacent points by an average;
\begin{align}
	\kappa_{i+1/2}^n = \frac{1}{2} (k_{i+1}^n + k_i^n) \\
	\kappa_{i-1/2}^n = \frac{1}{2} (k_{i-1}^n + k_i^n)
\end{align}
where $k_i^n$ is the thermal conductivity of the material at location $x_i$ at time $t_n$.

The discretised Eq. \ref{eq:heatEq_withSource}, with $V_i^n = V_\mathrm{meas}(x_i, t_n)$, then becomes;
\begin{equation}
	\begin{split}
		&C_i \rho_i \frac{T_i^{n+1} - T_i^n}{\Delta t} = \\ 
		&\frac{1}{2\Delta x^2} \left[ \kappa_{i+1/2}\left( T_{i+1}^{n+1} - T_{i}^{n+1} \right) - \kappa_{i-1/2}\left( T_{i}^{n+1} - T_{i-1}^{n+1} \right) \right]  \\
		&+ \frac{1}{2\Delta x^2} \left[ \kappa_{i+1/2}\left( T_{i+1}^{n} - T_{i}^{n} \right) - \kappa_{i-1/2}\left( T_{i}^{n} - T_{i-1}^{n} \right) \right]  \\
		&+ \frac{(V_i^n)^2 + (V_i^{n+1})^2}{2R_\text{s}v_\text{s}}
	\end{split}
\end{equation}

When obtaining the steady state surface temperature discussed in Section \ref{sec:steadyStateResults}, the Neumann condition from Equation \ref{eq:Neumann} is replaced by a Dirichlet condition where the boundary temperature is obtained by refining guesses that minimise the difference between simulated sensor temperature and measured sensor temperature.
		
	\printbibliography
	
\end{document}